\documentclass[pra,
 reprint,
 amsmath,amssymb,
 aps,
 nofootinbib
]{revtex4-2}

\usepackage{graphicx}
\usepackage{dcolumn}
\usepackage{bm}

\usepackage[utf8]{inputenc}
\usepackage[english]{babel}
\usepackage[dvipsnames]{xcolor}
\usepackage{hyperref}

\usepackage{amsmath,amssymb}
\usepackage{mathtools}
\usepackage{mathrsfs}
\usepackage{physics}
\usepackage{dsfont}
\usepackage{svg}
\usepackage[percent]{overpic} 

\usepackage{soul}

\usepackage{tikz}
\usetikzlibrary{arrows.meta, positioning}
\usetikzlibrary{quantikz2}

\usepackage[percent]{overpic} 

\usepackage{fancyhdr}

\usepackage[caption=false]{subfig} 

\ExplSyntaxOn
\makeatletter
\newsavebox{\@bra}
\newsavebox{\@brb}
\DeclarePairedDelimiterX\myinnerp[1]{.}{.}{
  \delimsize\langle
  \hspace*{0.3mm}\hspace*{0.55mm}
  \savebox{\@bra}{$\left\langle\vphantom{#1}\right.$}
  \hspace*{-0.9\wd\@bra}
  \delimsize\langle
  #1
  \delimsize\rangle
  \hspace*{0.3mm}\hspace*{0.55mm}
  \savebox{\@brb}{$\left.\vphantom{#1}\right\rangle$}
  \hspace*{-0.9\wd\@brb}
  \delimsize\rangle
}
\makeatother
\ExplSyntaxOff

\begin{document}


\title{Generating uniform quantum state ensembles with continuous measurement}

\author{Theodore McKeever}
\author{Ahsan Nazir}
\author{Harry J. D. Miller}

\affiliation{Department of Physics and Astronomy, The University of Manchester, Manchester I3 9PL, UK.}
\date{\today}

\begin{abstract}

We investigate the generation of uniform quantum state ensembles via continuous measurement. Using the $SU(d)$ Bloch representation, we derive the associated Langevin and Fokker--Planck equations and identify geometric conditions under which homogeneous monitoring causes global convergence to the uniform pure-state ensemble.
We then extend the analysis to mixed states, showing that homogeneous purity-dependent decoherence rates generate uniform Hilbert–Schmidt and Bures ensembles of qubit states through an effective nonlinear stochastic evolution.
Additionally, we introduce a post-mixing protocol for qubits: target mixed-state ensembles are assembled by classically sampling trajectories generated with different fixed efficiencies (or decoherence rates). This provides an experimentally feasible route to reconstructing Hilbert--Schmidt and Bures-random mixed-state ensembles, demonstrating that continuous monitoring provides both an exact dynamical generator of Haar-random pure states and a practical route to constructing mixed-state ensembles.

\end{abstract}

\maketitle


\section{Introduction}

The generation of random quantum states is a fundamental task across quantum science. In many-body systems, Haar-random pure states exhibit concentration of measure and near-maximal entanglement, underpinning typicality arguments and the emergence of thermal behaviour in subsystems \cite{Page1993,Popescu2006,Goldstein2006}. More generally, random states provide a natural framework for analysing generic properties of quantum channels and protocols, enabling average-case characterisation of quantities such as entropies and fidelities \cite{NielsenChuang}. Random density matrices, sampled according to measures such as the Hilbert--Schmidt (HS) and Bures metrics, are widely used to probe quantum state-space geometry and model generic noisy states \cite{Zyczkowski2001,BengtssonZyczkowski}. Standard constructions typically rely on random unitaries \cite{Emerson2005,Mezzadri2007} or dilation-based sampling methods \cite{Zyczkowski2001,CollinsNechita2016}.
Recent work has also shown that random quantum states can emerge from measurement-assisted chaotic dynamics \cite{Soonwon2022,Choi2023,Zhang2025}. By contrast, here we ask whether continuous weak measurement alone can be engineered to generate prescribed quantum state ensembles.

Stochastic master equations (SMEs) provide a natural framework for describing continuously monitored quantum systems, combining measurement back-action, decoherence, and Hamiltonian dynamics. While individual trajectories are stochastic, their ensemble defines a probability distribution over state space governed by a Fokker--Planck equation (FPE) \cite{HJDMCovariant2025}. This raises a central question: can continuous monitoring dynamically generate uniform quantum ensembles? A number of works have explored related aspects of measurement-induced stochastic dynamics \cite{Jackson2019ContinuousCoherent,Benoist2019InvariantSchrodinger,liu2025measurementbasedquantumdiffusionmodels,Shojaee018MeasurementTomography,Jacobs2016ContinuousMeasurement}. However, a direct geometric characterisation of when continuous monitoring generates uniform state ensembles remains incomplete.


In this work, we address this question by analysing the Bloch-space Langevin and Fokker--Planck dynamics associated with continuously monitored systems. We first focus on $d$-dimensional pure states, where the target distribution is geometrically natural: the uniform Fubini--Study (FS) measure on $\mathbb{CP}^{d-1}$. We derive geometric conditions under which this distribution is generated and show that they are satisfied by homogeneous monitoring (equal measurement strengths and unit efficiency). In this regime, the dynamics drives states to the pure-state manifold and produces the uniform ensemble, independent of the initial condition.

We then turn to mixed states, where the problem is more constrained. Existing production methods involve experimentally-challenging partial trace operations of large-dimensional random pure states \cite{RandomDensity2011}. In an effort to avoid such operations, we propose two new procedures founded on the dynamics of continuous measurement. First, we show that allowing the decoherence rate to depend on the purity coordinate, via non-Markovian feedback control \cite{Wiseman_Milburn_2009}, yields a closed stationarity condition for uniform bulk distributions. In the HS and Bures geometries, this admits regular and non-negative solutions for qubits. 

Second, motivated by the potential experimental difficulty in implementing state-dependent rates, we introduce another approach based on tunable measurement efficiencies. For qubits, homogeneous monitoring with different fixed efficiencies generates a family of isotropic radial distributions. Target ensembles can then be assembled through classical post-mixing of trajectories with different efficiencies---the same can be said for decoherence rates. This provides a practical route to approximating specified mixed state ensembles without requiring nonlinear state-dependent control.

From an operational perspective, these results establish continuous measurement setups as settings for ensemble generation: providing an exact route to Haar-random pure states, and options for potentially flexible and experimentally accessible mechanisms for mixed-state ensembles. This suggests applications in quantum state preparation, benchmarking, and noise characterisation in near-term quantum devices.

The paper is organised as follows. In Sec.~\ref{sec:SME} we introduce the monitored SME and derive the Bloch-space Langevin and Fokker--Planck equations in Sec.~\ref{sec:FPE}. In Sec.~\ref{sec:PureStateConditions} we formulate conditions for uniform pure-state generation, which are verified for homogeneous monitoring in Sec.~\ref{sec:PureStateProofs}. In Sec.~\ref{sec:MixedStates} and Sec.~\ref{sec:MixedStateProofs} we analyse the mixed-state problem and feedback-controlled rates solution. In Sec.~\ref{sec:inefficient} we then introduce the post-mixing approach. We conclude with a discussion in Sec.~\ref{sec:Discussion}.

\section{Bloch-space dynamics}
\label{sec:SME}

We consider a general $d$-level quantum system continuously monitored in a complete Hermitian operator basis $\{S_j\}_{j=1}^{d^2-1}$, with $\Tr(S_i S_j)=2\delta_{ij}$ and SU(d) Lie algebra defined by \cite{pfeifer2003lie}
\begin{align}
    [S_i,S_j]= i f_{ijk} S_k,
    \qquad
    \{S_i,S_j\}=\tfrac{4}{d}\delta_{ij}\mathds{1}+2 g_{ijk}S_k.
    \label{eq:SUd_algebra}
\end{align}
where Einstein summation notation is assumed. Here, $f_{ijk}$ and $g_{ijk}$ are, respectively, the real-valued antisymmetric and symmetric $SU(d)$ structure constants. We work throughout in the Bloch representation \cite{KIMURA2003339}
\begin{align}
    \rho(\vec{x})=\tfrac{\mathds{1}}{d}+\tfrac{1}{2}x_j S_j,
    \qquad
    x_j=\Tr(S_j\rho),
    \label{eq:bloch_rep}
\end{align}
and take the Hamiltonian to be traceless, $H=h_j S_j$.

Later on, our aim will be to discern constraints on state dynamics such that uniform state distributions are generated. Therefore, we begin with a general weak-measurement SME with additional decoherence channels.
The monitored stochastic master equation is \cite{ALBARELLI2024129260}
\begin{align}
    d\rho
    =&
    -i[H,\rho]\,dt
    +\sum_j \mathcal{D}[\xi_j]\rho\,dt \nonumber \\
    & \ \ \ 
    +\sum_j \mathcal{D}[c_j]\rho\,dt
    +\sum_j \sqrt{\eta_j}\,\mathcal{H}[c_j]\rho\,dW_j,
    \label{eq:SME_main}
\end{align}
with independent Wiener increments satisfying $dW_i\,dW_j=\delta_{ij}\,dt$ \cite{gardiner2009stochastic}. We restrict our attention to the measurement of Hermitian observables, and also work in dimensionless units here and throughout. The measurement and decoherence jump operators are thus represented by
$c_j=\sqrt{2k_j}\,S_j$ and $\xi_j=\sqrt{2\gamma_j}\,S_j$, respectively, so it is convenient to define the joint rate
\begin{align}
    l_j:=2k_j+2\gamma_j.
\end{align}
The Lindblad and stochastic innovation superoperators for Hermitian observables $c$ are
\begin{align}
    \mathcal{D}[c]\rho & =c\rho c-\tfrac{1}{2}\{c^2,\rho\}
\end{align}
and
\begin{align}
    \mathcal{H}[c]\rho& =c\rho+\rho c-2\Tr(c\rho)\rho .
\end{align}
In Eq.~\eqref{eq:SME_main}, we purposefully allow for simultaneous measurement of non-commuting observables, which is valid under sufficiently weak measurements (for discussion, see App.~\ref{app:weakMeasurementCompatibility}).

Expanding Eq.~\eqref{eq:SME_main} in the Bloch basis using $dx_l=\Tr(S_l\,d\rho)$ yields a Langevin equation for the Bloch components,
\begin{align}
    dx_l=f^l(\vec{x})\,dt+\sigma^l_j(\vec{x})\,dW_j,
    \label{eq:langevin_compact}
\end{align}
where $f^l$ is the drift vector and $\sigma^l_j$ are noise amplitudes which, in general, depend on the state $\vec{x} = (x_1,...,x_{d^2-1})^T$.
The full derivation is algebraic but straightforward; we defer details to App.~\ref{app:LangevinDerivation} and quote the result here.

First, the Hamiltonian contribution is purely rotational:
\begin{align}
    f^l_{H}=\sum_{j,k} h_j f_{ljk}x_k.
    \label{eq:hamiltonian_drift}
\end{align}
Second, the dissipative contribution is linear in $\vec{x}$ and may be written as
\begin{align}
    f^l_{\mathrm{diss}}
    =&
    \tfrac{2}{d}\sum_j l_j g_{ljj}
    +\tfrac{2}{d}x_l\!\Big(l_l-\sum_j l_j\Big) \nonumber \\
    &  +\sum_{j,m,k} l_j
    \Big[
        \left(\tfrac{i}{2}f_{jml}+g_{jml}\right)
        \left(\tfrac{i}{2}f_{jmk}+g_{jmk}\right) \nonumber \\
    & \ \ \ \ \ \ \ \ \ \ \ \ \ \ \ \ \ \ \ \ \ \ \ \ \ \ \ \ \ \ \ \ \ \    -g_{jjm}g_{lkm}
    \Big]x_k.
    \label{eq:dissipative_drift}
\end{align}
Thus the total drift is
\begin{align}
    f^l=f^l_{H}+f^l_{\mathrm{diss}}.
    \label{eq:drift_total}
\end{align}

Finally, the noise amplitudes are equal to 
\begin{align}
    \sigma^l_j(\vec{x})
    =
    \sqrt{2k_j\eta_j}
    \left(
        \tfrac{4}{d}\delta_{lj}
        +2g_{ljm}x_m
        -2x_l x_j
    \right).
    \label{eq:noise_amplitudes_main}
\end{align}
From these amplitudes, the diffusion tensor
\begin{align}
    D^{\mu\nu}=\sum_j \sigma^\mu_j \sigma^\nu_j
    \label{eq:diffusion_tensor_def}
\end{align}
may be written explicitly as
\begin{widetext}
\begin{align}
    D^{\mu\nu}=  \sum_j 2 k_j \eta_j \bigg[&
    \tfrac{16}{d^2} \delta_{\mu j}\delta_{\nu j}
    + \delta_{\mu j}\tfrac{8}{d}\sum_n g_{\nu j n}x_n
    - \delta_{\mu j}\tfrac{8}{d}x_\nu x_j
    + \delta_{\nu j}\tfrac{8}{d}\sum_m g_{\mu j m}x_m
    \nonumber\\
    &\hspace{0.2cm}
    + 4\sum_{m,n} g_{\mu jm}g_{\nu jn}x_n x_m
    - 4\sum_m g_{\mu jm}x_m x_\nu x_j
    - \delta_{\nu j}\tfrac{8}{d}x_\mu x_j
    - 4\sum_n g_{\nu jn}x_\mu x_j x_n
    + 4 x_\mu x_j x_\nu x_j
    \bigg].
    \label{eq:Duv}
\end{align}
\end{widetext}
This form will be useful when extracting particular instances of the diffusion tensor and its derivatives. For simplifications and derivative expressions of $D^{\mu\nu}$ used throughout, see App.~\ref{app:FPE_terms}.

To aid such simplifications, several standard contractions of the $SU(d)$ structure constants will be used repeatedly \cite{Georgi2000}:
\begin{align}
    &\sum_{a,b} f_{abi}f_{abj}=4d\,\delta_{ij},
    \qquad
    \sum_{a,b} f_{abi}g_{abj}=0,
    \qquad \nonumber \\
    &\sum_{a,b} g_{abi}g_{abj}=\frac{d^2-4}{d}\,\delta_{ij},
    \qquad
    \sum_a g_{aab}=0.
    \label{eq:structure_identities}
\end{align}
These enter when simplifying both the drift \eqref{eq:dissipative_drift} and the diffusion tensor \eqref{eq:Duv}.

\section{Covariant Fokker--Planck equation}
\label{sec:FPE}

Equation \eqref{eq:langevin_compact} defines a diffusion process on the Bloch-coordinate manifold. To describe the evolution of an ensemble of trajectories, one passes from the stochastic differential equation for $\vec{x}_t$ to a FPE for the associated probability density. Since we ultimately wish to discuss geometry-dependent notions of uniformity, and are concerned with mixed state evolution for which there is no unique distance measure, it is useful to state this equation in covariant form from the outset.

Let $\varrho(\vec{x},t)$ denote the geometric probability density with respect to a chosen metric $g_{\mu\nu}$ on the state space manifold, $\mathcal M$. The choice of metric $g_{\mu\nu}$ informs the covariant derivative, $\nabla_\mu$. In the HS geometry, 
the Bloch coordinates are Euclidean where $g_{\mu\nu}=\delta_{\mu\nu}$ and $\nabla_\mu = \partial_\mu$, but we keep the notation covariant here because later sections move beyond flat geometry. 

The covariant FPE associated with Eq.~\eqref{eq:langevin_compact} is
\begin{align}
    \partial_t\varrho=-\nabla_\nu J^\nu
    \label{eq:FPE_covariant}
\end{align}
in the conservation-law form, where the probability current is
\begin{align}
    J^\nu:=h^\nu \varrho-\tfrac{1}{2}\nabla_\mu(D^{\mu\nu}\varrho).
    \label{eq:probability_current}
\end{align}
The drift vector from the Langevin equation inherits a covariant correction,
\begin{align}
    h^\nu = f^\nu + \tfrac12 \Gamma_{ij}^\nu D^{ij},
    \label{eq:hv}
\end{align}
where $f^\nu$ is the deterministic drift, $\tfrac12 \Gamma_{ij}^\nu D^{ij}$ is the geometric drift and $\Gamma_{ij}^\nu$ are the Christoffel symbols associated with the chosen metric \cite{GRAHAM1985209}. Both $h^\nu$ and the diffusion tensor are functions of the state.

The passage from the Langevin equation \eqref{eq:langevin_compact} to the covariant FPE \eqref{eq:FPE_covariant} follows from a standard It\^o expansion of test functions together with the requirement of coordinate invariance; we defer the full derivation to App.~\ref{app:CovariantFPE}. For present purposes, the key point is that the drift $h^\nu$ determines deterministic transport on state space, while the quadratic variation of the noise amplitudes produces the second-order diffusive term through the tensor $D^{\mu\nu}$ defined in Eq.~\eqref{eq:diffusion_tensor_def}.

Associated with Eq.~\eqref{eq:FPE_covariant} is the backward Kolmogorov generator $L$, defined as the adjoint of the forward operator $L^*$ \cite{risken1989fpe}:
\begin{align}
    \int_{\mathcal M} y\,(L^*\varrho)\,dV
    =
    \int_{\mathcal M} (Ly)\,\varrho\,dV,
    \label{eq:adjoint_relation}
\end{align}
for all sufficiently regular scalar test functions $y$. For Eq.~\eqref{eq:langevin_compact}, the generator is
\begin{align}
    Ly
    =
    h^\nu \partial_\nu y
    +\tfrac{1}{2}D^{\mu\nu}\nabla_\mu\partial_\nu y.
    \label{eq:backward_generator}
\end{align}
Since $y$ is a scalar, the first derivative is simply $\nabla_\nu y=\partial_\nu y$.

Stationary distributions $\varrho_*$ satisfy $L^*\varrho_*=0$, or equivalently $\nabla_\nu J^\nu=0$. A stronger condition, corresponding to detailed balance, is the vanishing of the current itself, $J^\nu=0$.

The expressions \eqref{eq:diffusion_tensor_def}, \eqref{eq:FPE_covariant} and \eqref{eq:hv} provide the starting point for the geometric analysis of uniform state ensembles. Before considering uniform ensembles over mixed states, in the next section, we restrict these dynamics to the pure-state manifold.

\section{Pure-State Uniformity}
\label{sec:PureStateConditions} 

\begin{figure*}[t]
    \centering
    
    
    \includegraphics[width=\textwidth]{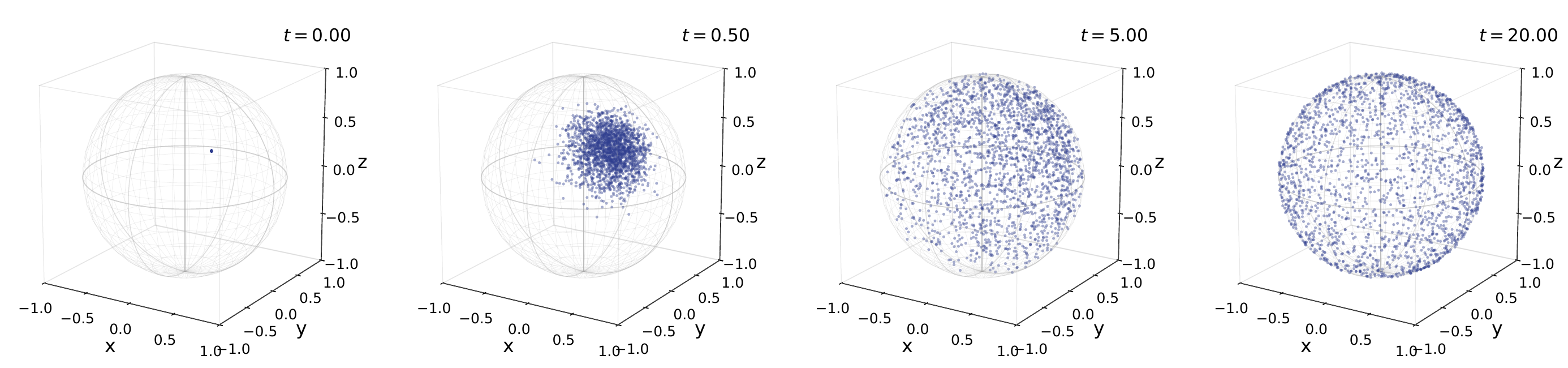}
    
    \vspace{0.0cm}
    
    \includegraphics[width=\textwidth]{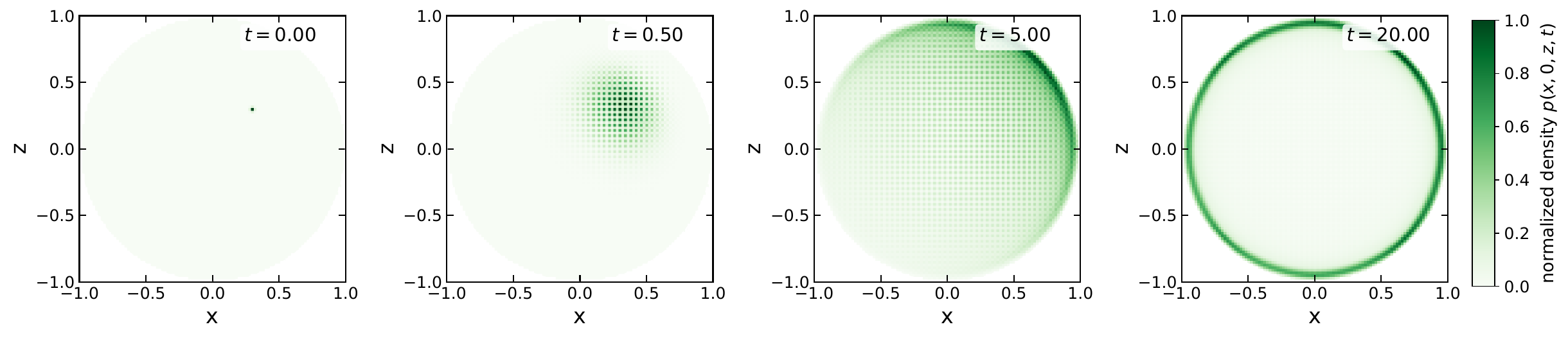}
    \caption{
    Ensemble dynamics in the Bloch ball for the homogeneous pure-state protocol with $\gamma=0$. 
    The top row shows the SME ensemble at selected times obtained from $2000$ individual trajectories, and the bottom row shows the associated $y=0$ cross-sections of the Fokker--Planck density in the $x$-$z$ plane. 
    In this case the dynamics drives the ensemble toward the pure-state boundary, where it relaxes to the uniform distribution over pure states. 
    The SME trajectories each start at $(x,y,z)=(0.3,0,0.3)$ and the FPE simulation is initialised as a narrow Gaussian centred at the same point with width $\sigma=0.01$.
    The Fokker--Planck simulations use a Cartesian grid with $N=101$ points per axis, timestep $dt=10^{-3}$, final time $t_{\mathrm{final}}=20$, vanishing Hamiltonian, and measurement strength $k=10^{-2}$.
    }
    \label{fig:pure_state_dynamics}
\end{figure*}

We now restrict attention to the generation and preservation of uniform ensembles of pure states. In contrast to the mixed-state problem, this target is geometrically unambiguous: the set of pure states of a $d$-level system forms the manifold $\mathbb{CP}^{d-1}$, equipped with the FS measure, which induces the natural notion of a uniform pure-state distribution. 

In the Bloch representation \eqref{eq:bloch_rep}, pure states are 
identified by imposing $\rho^2=\rho$. As laid out in App.~\ref{app:PureStateIdentities}, matching identity and generator components yields 
\begin{align}
    x_a x_a = \frac{2(d-1)}{d},
    \qquad
    \sum_{a,b} g_{abn}x_a x_b = \frac{2(d-2)}{d}x_n.
    \label{eq:pure_state_identities}
\end{align}
Thus, pure states obey the relations in Eq.~\eqref{eq:pure_state_identities} and can be identified using the purity coordinate 
\begin{align}
    C(x):=\Tr(\rho^2)-1
    =\frac{1}{2}\sum_a x_a^2-\frac{d-1}{d}.
    \label{eq:purity_constraint}
\end{align}
Hence $C(x)=0$ defines a feature of the pure-state manifold, while $\frac{1-d}{d}\leq C(x)<0$ corresponds to the mixed-state interior. The gradient provides the outward normal direction in the embedding space $\mathbb{R}^{d^2-1}$. We therefore define the unnormalised normal vector
\begin{align}
    n_\nu := \partial_\nu C = x_\nu.
    \label{eq:normal_vector}
\end{align}

This allows any vector field to be decomposed into normal and tangential components relative to the level set $C=0$. The corresponding tangential projector is
\begin{align}
    P^\mu_\nu
    =
    \delta^\mu_\nu-\frac{x^\mu x_\nu}{|x|^2},
    \label{eq:tangent_projector}
\end{align}
where $|x|^2=x_\lambda x_\lambda$. This definition follows from decomposing any vector $v^\mu$ into normal and tangential components relative to $n_\nu$:
\begin{align}
    v^\mu = (v \cdot n)\frac{n^\mu}{|n|^2} + \left(v^\mu - (v \cdot n)\frac{n^\mu}{|n|^2}\right),
\end{align}
so that the tangential component, $\tilde{v}$ is given by the second term.

Since $P^2=P$, any tangent vector $\tilde v^\mu:=P^\mu_\nu v^\nu$ is invariant under further projection. We use this projector to define tangential components of the drift and diffusion:
\begin{align}
    \tilde h^\mu := P^\mu_\nu h^\nu,
    \qquad
    \tilde D^{\mu\nu}:=P^\mu_\alpha D^{\alpha\beta}P^\nu_\beta.
    \label{eq:projected_fd}
\end{align}
Likewise, projections can be applied to derivatives of $h$ and $D$.



\subsection*{Conditions for pure-state uniformity}

Now consider the Fokker--Planck dynamics of Eq.~\eqref{eq:FPE_covariant} induced by the monitored SME in Eq.~\eqref{eq:SME_main}. A sufficient (but not necessarily unique) set of conditions for the uniform distribution on the pure-state manifold $C=0$ to be uniquely generated and preserved is:
\begin{enumerate}
    \item[\textbf{(i)}] \textbf{No leakage off the pure-state manifold:}
    \begin{align}
        h^\nu \partial_\nu C
        =
        \tfrac{1}{2}(\nabla_\mu D^{\mu\nu})\partial_\nu C
        \qquad \text{on } C=0.
        \tag{i}
        \label{eq:cond_i}
    \end{align}
    A stronger condition is
    \begin{align}
        D^{\mu\nu}\partial_\nu C=0
        \qquad \forall \mu
        \qquad \text{on } C=0,
        \tag{i$^*$}
        \label{eq:cond_i_strong}
    \end{align}
    which enforces strictly tangential diffusion at the boundary.

    \item[\textbf{(ii)}] \textbf{Flow toward the pure-state manifold:}
    \begin{align}
        LC>0
        \qquad \text{for } C<0,
        \tag{ii}
        \label{eq:cond_ii}
    \end{align}
    where $L$ is the backward generator from Eq.~\eqref{eq:backward_generator}.

    \item[\textbf{(iii)}] \textbf{Stationarity of the uniform distribution:}
    \begin{align}
        -\tilde{\nabla}_\mu \tilde h^\mu
        +
        \tfrac{1}{2}\tilde{\nabla}_\mu \tilde{\nabla}_\nu \tilde D^{\mu\nu}
        =0
        \qquad \text{on } C=0,
        \tag{iii}
        \label{eq:cond_iii}
    \end{align}
    where, together with \eqref{eq:cond_i}, we are interested only in the tangential dynamics.
    A stronger, detailed-balance condition is
    \begin{align}
        \tilde h^\mu - \tfrac{1}{2}\tilde{\nabla}_\nu \tilde D^{\nu\mu}=0 
        \qquad \forall \mu \text{ on } C=0,
        \tag{iii$^*$}
        \label{eq:cond_iii_strong}
    \end{align}

    \item[\textbf{(iv)}] \textbf{Isotropic diffusion:}
    \begin{align}
        \tilde v_\mu \tilde D^{\mu\nu}\tilde v_\nu >0
        \qquad
        \text{for all } \tilde v^\mu \neq 0,
        \tag{iv}
        \label{eq:cond_iv}
    \end{align}
    with $\tilde v^\mu \partial_\mu C=0$.
\end{enumerate}

\medskip

The intuition behind these conditions is as follows. Condition~\eqref{eq:cond_i} ensures that probability supported on $C=0$ is not driven off the pure-state manifold by a net normal current. It arises by imposing $J^\nu =0$ for a uniform distribution $\varrho_*$ on $C=0$, noting that $\partial_\mu \varrho_*=0$, and contracting the resulting expression with the normal vector $n_\nu=\partial_\nu C$. Condition~\eqref{eq:cond_ii} guarantees attraction from the mixed-state bulk toward the pure shell. Condition~\eqref{eq:cond_iii} ensures that, once restricted to the pure-state manifold, the induced tangential dynamics preserves the uniform distribution. Finally, condition~\eqref{eq:cond_iv} ensures that the tangential diffusion is non-degenerate (non-singular), so that all directions along the pure-state manifold are explored; together with \eqref{eq:cond_i}--\eqref{eq:cond_iii}, this yields irreducibility and hence uniqueness of the steady state.

The set \eqref{eq:cond_i}--\eqref{eq:cond_iv} is sufficient for global generation of the uniform pure-state ensemble, while \eqref{eq:cond_i_strong} and \eqref{eq:cond_iii_strong} impose stricter geometric structure, namely strictly tangential diffusion and detailed balance on the pure shell.

In the next section, we show that these conditions are satisfied for \emph{homogeneous} continuous monitoring, that is, complete symmetry over measurement channels.

\begin{figure}[t]
    \centering
    \includegraphics[width=0.95\columnwidth]{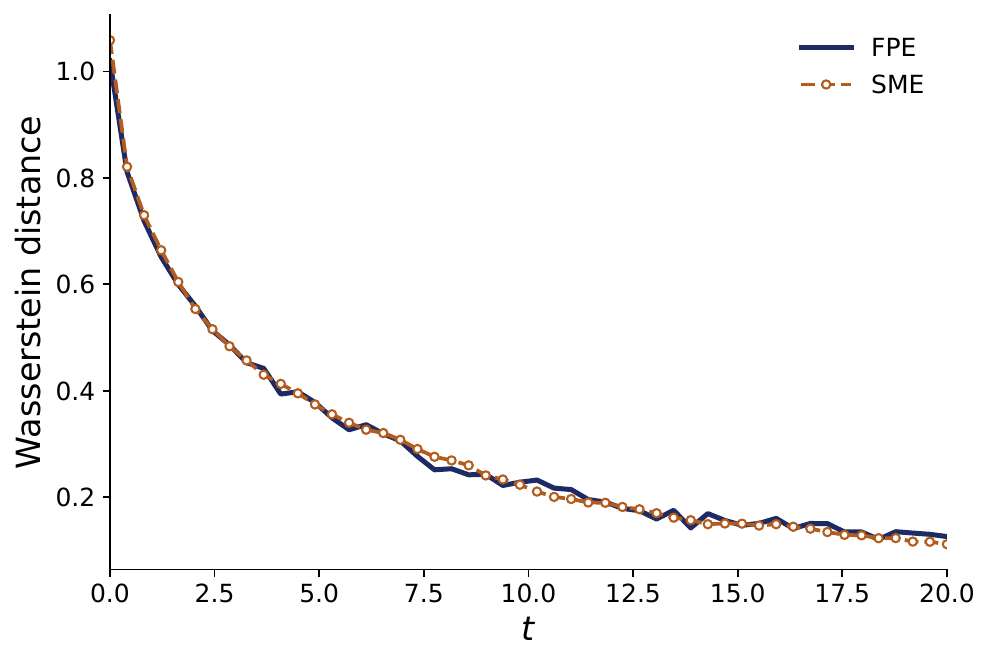}
    \caption{
    Wasserstein-1 distance (with entropic regularisation) to the uniform pure-state reference distribution for the qubit homogeneous monitoring protocol with $\gamma=0$. 
    The solid curve is obtained using $2000$ samples from the evolving Fokker--Planck density, while the dashed curve with markers is obtained from the same number of corresponding SME trajectories.
    Both curves demonstrate monotonic convergence toward the uniform pure-state ensemble, with finite-sample fluctuations.
    }
    \label{fig:pure_state_wasserstein}
\end{figure}


\section{Homogeneous monitoring}
\label{sec:PureStateProofs}

We now verify that the conditions \eqref{eq:cond_i}--\eqref{eq:cond_iv} of Sec.~\ref{sec:PureStateConditions} are satisfied for homogeneous continuous monitoring. Specifically, we consider the symmetric setting
\begin{align}
    k_j = k,
    \qquad
    \eta_j = 1,
    \qquad
    \gamma_j = 0,
    \label{eq:symmetric_conditions}
\end{align}
corresponding to equal measurement strengths, unit efficiency, and no additional decoherence. Under this setup the SME is dynamically equivalent to stochastic Schrödinger descriptions when the initial state is pure.

In the analysis below, the dynamics are first expressed in the ambient Bloch coordinates equipped with the flat HS metric and are then restricted to the invariant pure-state manifold through the constraint $C=0$. In these ambient Cartesian coordinates the Christoffel symbols vanish, so the effective drift appearing in Eq.~\eqref{eq:hv} coincides with the deterministic drift, $h^\mu=f^\mu$, for the purposes of this section. Accordingly, all covariant quantities appearing in conditions~\eqref{eq:cond_i}--\eqref{eq:cond_iv} are evaluated in the ambient HS coordinates; in particular, the covariant derivative $\nabla_\nu$ reduces to the ordinary partial derivative $\partial_\nu$.

Under homogeneous conditions, the drift and diffusion simplify considerably. Using the identities \eqref{eq:structure_identities} and restricting to the pure-state shell using the relations \eqref{eq:pure_state_identities}, one finds that\footnote{See App.~\ref{app:FPE_terms}.}
\begin{align}
    f^\nu &= \sum_{j,k} h_j f_{\nu jk} x_k - 4kd\, x_\nu,
    \label{eq:f_simplified}
\end{align}
and
\begin{align}\label{eq:homo_Duv}
    D^{\mu\nu}
    =
    & 2k
    \Big[
        \delta_{\mu \nu} \frac{16}{d^2}
        + \frac{16}{d} \sum_n g_{\nu \mu n}x_n \nonumber\\
        &+ 4\sum_{j,m,n} g_{\mu jm} g_{\nu jn}x_n x_m
        - 8\left( 1-\tfrac{1}{d}\right) x_\mu x_\nu
    \Big].
\end{align}
To evaluate the divergence of the diffusion tensor, one must differentiate the homogeneous form of Eq.~\eqref{eq:Duv} before imposing the pure-state identities, resulting in the expression,
\begin{align}\label{eq:divD_simplified}
    \partial_\mu D^{\mu\nu} = -8kd\,x_\nu.
\end{align}

We now verify each condition in turn.

\subsection*{(i) No leakage off the pure-state manifold}

We first verify condition \eqref{eq:cond_i}. Using Eqs.~\eqref{eq:normal_vector} and \eqref{eq:f_simplified}, we compute
\begin{align}
    f^\nu \partial_\nu C
    = \sum_{j,k} h_j f_{\nu jk} x_k x_\nu - 4kd\, x_\nu x_\nu. 
\end{align}
The first term vanishes by antisymmetry of $f_{\nu jk}$ in $\nu,k$, so
\begin{align} \label{eq:(i)fdC}
    f^\nu \partial_\nu C = -4kd\, |x|^2.
\end{align}
Similarly from Eq.~\eqref{eq:divD_simplified},
\begin{align}\label{eq:(i)dDdC}
    \tfrac{1}{2}(\partial_\mu D^{\mu\nu})\partial_\nu C
    = \tfrac{1}{2}(-8kd\, x_\nu)x_\nu
    = -4kd\, |x|^2.
\end{align}
Thus condition \eqref{eq:cond_i} is satisfied straightforwardly. Note that, if $\gamma \neq 0$, the replacement  $k\rightarrow k+\gamma$ would be required in Eq.~\eqref{eq:(i)fdC} and, if $\eta \neq 1$, $k\rightarrow k \eta$ in Eq.~\eqref{eq:(i)dDdC}; in either case \eqref{eq:cond_i} would not hold.

\medskip

We now outline the verification of the stronger condition \eqref{eq:cond_i_strong}, namely that diffusion is strictly tangential on $C=0$. 
Contracting Eq.~\eqref{eq:homo_Duv} with $x_\nu$, and using the identities \eqref{eq:pure_state_identities}, one finds that all terms cancel pairwise on $C=0$, yielding $ D^{\mu\nu}x_\nu = 0$.
Thus the diffusion has no component in the normal direction, and is entirely tangent to the pure-state manifold. This establishes condition \eqref{eq:cond_i_strong}.

\subsection*{(ii) Flow toward the pure-state manifold}

The most direct way to establish condition~\eqref{eq:cond_ii} is to work in operator space, returning to the stochastic master equation \eqref{eq:SME_main}. In this setting, the backward generator $L$ acts on functions of the state, $y(\rho)$, according to \cite{risken1989fpe}
\begin{align}
    L y(\rho)
    =
    \lim_{dt \to 0}
    \frac{\langle \!\langle y(\rho_{t+dt}) - y(\rho_t) \rangle\!\rangle}{dt}.
    \label{eq:Lderivative}
\end{align}
where $\langle \!\langle...\rangle \!\rangle$ denotes an expectation over noise realisations.

To evaluate $LC$, we apply It\^o calculus to the purity function $C=\Tr(\rho^2)-1$. Using the matrix It\^o rule $d(\rho^2) = (d\rho)\rho + \rho(d\rho) + (d\rho)^2$,
we obtain
\begin{align}
    dC = d\Tr(\rho^2)
    =
    2\Tr(\rho\,d\rho)
    +
    \Tr\!\big((d\rho)^2\big).
\end{align}

Writing the SME \eqref{eq:SME_main} in the form
\begin{align}\label{eq:SME_ABform}
    d\rho = A\,dt + \sum_j B_j\,dW_j,
\end{align}
with independent Wiener increments satisfying $dW_i dW_j = \delta_{ij}dt$, we have
\begin{align}
    (d\rho)^2 = \sum_j B_j^2\,dt
\end{align}
and thus
\begin{align}
    dC
    =
    \Big(
        2\Tr(A\rho)
        +
        \sum_j \Tr(B_j^2)
    \Big)dt
    +
    2\sum_j \Tr(B_j\rho)\,dW_j.
\end{align}

Substituting into Eq.~\eqref{eq:Lderivative}, the stochastic term averages to zero, yielding
\begin{align}
    LC = 2\Tr(A\rho) + \sum_j \Tr(B_j^2).
\end{align}
Using the explicit SME \eqref{eq:SME_main}, this becomes
\begin{align}\label{eq:(ii)SME}
    LC
    =&
    -2i\Tr\!\big(\rho[H,\rho]\big)
    +
    2 \sum_j l_j \Tr\!\big(\rho\,\mathcal{D}[S_j]\rho\big)
    \nonumber \\
    &\ \ \ \ \ \ +
    \sum_j 2k_j\eta_j \Tr\!\big((\mathcal{H}[S_j]\rho)^2\big).
\end{align}

The Hamiltonian term vanishes by cyclicity of the trace and the dissipative contribution evaluates to
\begin{align}
    2\Tr\!\big(\rho\,\mathcal{D}[S_j]\rho\big)
    =
    \Tr\!\big([S_j,\rho]^2\big).
\end{align}
Defining $F_j := S_j - \Tr(S_j\rho)$ so that $\mathcal{H}[S_j]\rho = F_j\rho + \rho F_j$, the stochastic term becomes
\begin{align}
    \Tr\!\big((\mathcal{H}[S_j]\rho)^2\big)
    =
    4\Tr(F_j\rho F_j\rho)
    -
    \Tr\!\big([S_j,\rho]^2\big).
\end{align}
For the pure-state protocol with $\gamma_j=0$ and $\eta_j=1$, these contributions combine in Eq.~\eqref{eq:(ii)SME} to give
\begin{align}
    LC = 8 \sum_j k_j \Tr(F_j \rho F_j \rho).
\end{align}

Lastly, to establish positivity, we define $X_j := \sqrt{\rho}\,F_j\,\sqrt{\rho}$, which is Hermitian since both $\rho$ and $F_j$ are Hermitian. The expression for $LC$ can then be expressed as
\begin{align}
    LC = 8 \sum_j k_j \Tr(X_j^2).
\end{align}
Since $\Tr(X_j^2)=\sum_a (\lambda_j^a)^2$, where $\lambda_j^a$ are the real eigenvalues of $X_j$, we obtain
\begin{align}
    LC = 8 \sum_{j,a} k_j (\lambda_j^a)^2 > 0
\end{align}
for all mixed states. This establishes condition~\eqref{eq:cond_ii}; the pure state manifold is universally attractive for mixed states in the absence of additional decoherence or inefficient measurements. Note that homogeneous measurement ($k_j=k$) is not required to satisfy \eqref{eq:cond_ii}.


\subsection*{(iii) Stationarity of the uniform distribution}

Straightforwardly, by combining Eqs.~\eqref{eq:f_simplified} and \eqref{eq:divD_simplified}, we find
\begin{align}
    f^\nu - \tfrac{1}{2}\partial_\mu D^{\mu\nu}
    =f_H^\nu.
\end{align}
Thus, on the pure-state manifold,
\begin{align}
    f^\nu - \tfrac{1}{2}\partial_\mu D^{\mu\nu}
    =
    \begin{cases}
        0, & H=0,\\
        \text{purely tangential}, & H\neq 0.
    \end{cases}
\end{align}
If $H=0$, the stronger condition \eqref{eq:cond_iii_strong} holds, and the system satisfies detailed balance. If $H\neq 0$, the remaining term generates Hamiltonian rotations which are divergence-free, since $\partial_\nu f_H^\nu =0$.

Thus condition \eqref{eq:cond_iii} is satisfied: the uniform distribution remains stationary, but with a non-vanishing tangential current corresponding to a non-equilibrium steady state.

\subsection*{(iv) Isotropic diffusion}

Condition \eqref{eq:cond_iv} concerns the projected tensor $\tilde D^{\mu\nu}=P^\mu_{\ \alpha}D^{\alpha\beta}P^\nu_{\ \beta}$. If $\tilde v^\mu$ is already tangent, then $P^\mu_{\ \nu}\tilde v^\nu=\tilde v^\mu$, and therefore
\begin{align}
    \tilde v_\mu \tilde D^{\mu\nu} \tilde v_\nu
    &=
    \tilde v_\mu P^\mu_{\ \alpha}D^{\alpha\beta}P^\nu_{\ \beta}\tilde v_\nu \nonumber\\
    &=
    \tilde v_\alpha D^{\alpha\beta}\tilde v_\beta.
\end{align}
Thus, when testing positivity on the tangent space, it is sufficient to contract $D^{\mu\nu}$ directly with tangent vectors.

From Eq.~\eqref{eq:diffusion_tensor_def}, we therefore obtain
\begin{align}
    \tilde v_\mu D^{\mu\nu}\tilde v_\nu
    =
    \sum_j (\tilde v_\mu \sigma^\mu_j)^2 \geq 0.
\end{align}
Hence the tangential diffusion is automatically positive semidefinite.

To prove strict positivity, we use the explicit form of the noise amplitudes \eqref{eq:noise_amplitudes_main} in the homogeneous case.
Contracting with a tangent vector $\tilde v^\mu$, and using $\tilde v^\mu x_\mu=0$, gives
\begin{align}
    \tilde v_\mu \sigma^\mu_j
    =
    2\sqrt{2k}\left(
        \tfrac{2}{d}\tilde v_j + g_{\mu j n}\tilde v_\mu x_n
    \right).
\end{align}
Therefore
\begin{align}
    \tilde v_\mu D^{\mu\nu}\tilde v_\nu
    =
    8k \sum_j
    \left(
        \tfrac{2}{d}\tilde v_j + g_{\mu j n}\tilde v_\mu x_n
    \right)^2.
    \label{eq:tangent_diffusion_squares}
\end{align}
This is a sum of squares, so it vanishes if and only if
\begin{align}
    \tfrac{2}{d}\tilde v_j + g_{\mu j n}\tilde v_\mu x_n = 0
    \qquad \forall j.
    \label{eq:kernel_condition}
\end{align}

On $C=0$, the conditions \eqref{eq:cond_i} and \eqref{eq:cond_i_strong} ensure that evolution and diffusion are purely tangential, so it suffices to consider only tangential evolution, $\dot{x}_a = \tilde{v}_a$. We now differentiate both sides of the pure-state identity \eqref{eq:pure_state_identities}, along a tangent direction $\tilde v$. This gives\footnote{See App.~\ref{app:PureStateIdentities}.}
\begin{align}\label{eq:tangent_g_identity}
    g_{abn}\tilde v_a x_b = \frac{(d-2)}{d}\tilde v_n.
\end{align}
Substituting Eq.~\eqref{eq:tangent_g_identity} into Eq.~\eqref{eq:kernel_condition} yields
\begin{align}
    \frac{2}{d}\tilde v_j + \frac{d-2}{d}\tilde v_j = \tilde v_j = 0
    \qquad \forall j.
\end{align}
Hence the only tangent vector for which Eq.~\eqref{eq:tangent_diffusion_squares} vanishes is the zero vector. Therefore $\tilde v_\mu \tilde D^{\mu\nu}\tilde v_\nu > 0$ for all nonzero tangent vectors $\tilde v^\mu$, and condition \eqref{eq:cond_iv} follows.

\medskip

\noindent
Taken together, conditions \eqref{eq:cond_i}--\eqref{eq:cond_iv} are satisfied for homogeneous continuous monitoring. We therefore conclude that the induced dynamics generates and preserves the uniform distribution over pure states for any (mixed or pure) initial state or distribution. In the absence of a Hamiltonian, this steady state satisfies detailed balance, while for $H\neq 0$ it becomes a non-equilibrium steady state with a divergence-free tangential probability current.

Numeric simulations of qubit SME trajectories and FPE densities confirm these conclusions, as presented in Fig.~\ref{fig:pure_state_dynamics}. The Wasserstein-1 distance measure\footnote{See App.~\ref{app:wasserstein_calc}.}, which quantifies the amount of effective work required to shape one normalised probability distribution into another \cite{Villani2009}, and is displayed in Fig.~\ref{fig:pure_state_wasserstein}, shows that qubits monotonically approach the uniform distribution when subjected to homogeneous measurement in the way described in this section.


\section{Mixed-state uniformity}
\label{sec:MixedStates}

\begin{figure*}[t]
    \centering
    
    
    \includegraphics[width=\textwidth]{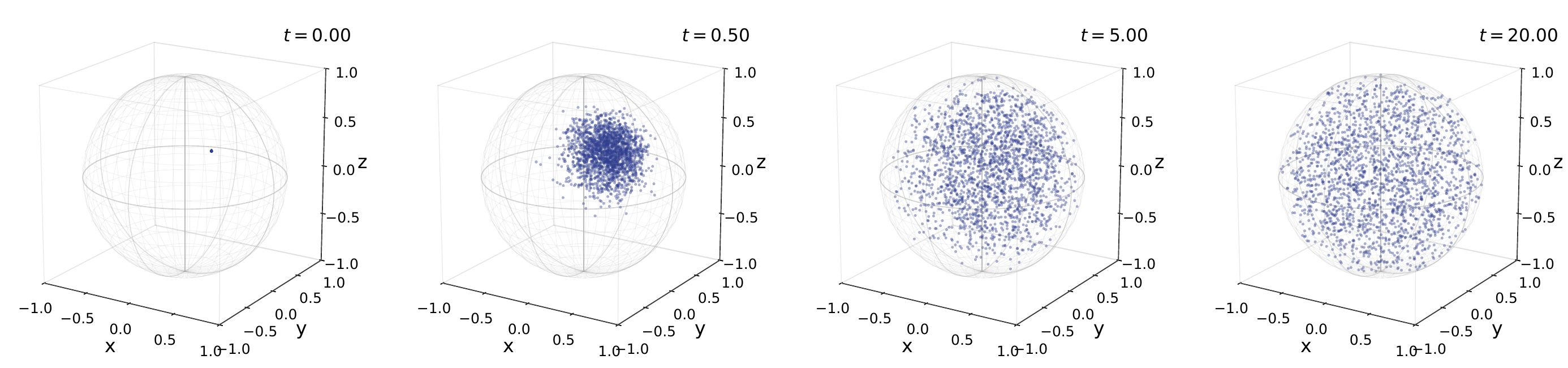}
    
    \vspace{0.0cm}
    
    \includegraphics[width=\textwidth]{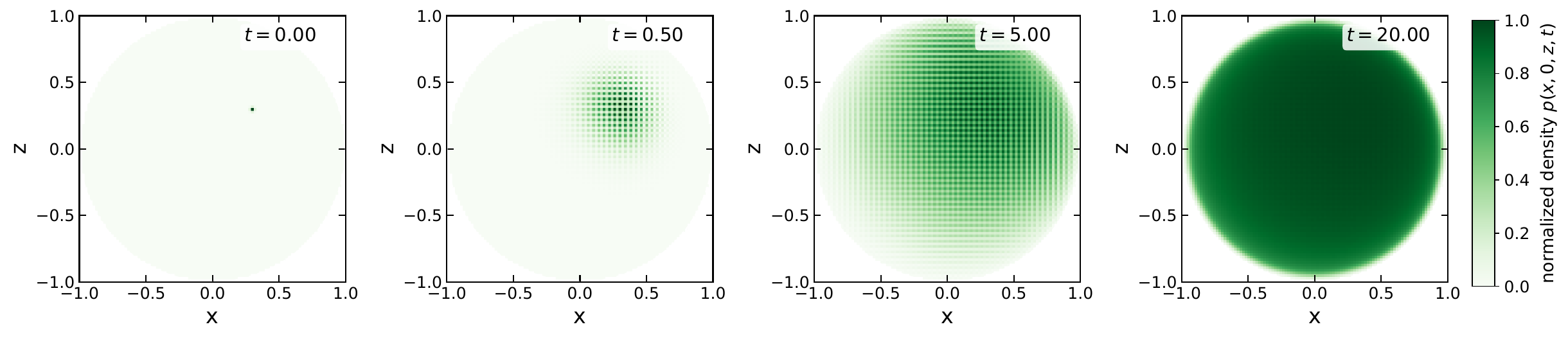}
    \caption{
    Ensemble dynamics in the Bloch ball for the state-dependent mixed-state protocol with $\gamma(C)=-6kC$, corresponding to the regular HS-uniform bulk solution for a qubit. 
    The top row shows the SME ensemble at selected times, rendered in the full Bloch ball and obtained from a finite set of trajectories. 
    The bottom row shows the associated $y=0$ cross-sections of the Fokker--Planck density in the $x$-$z$ plane.
    In each case, we see that the dynamics spreads through the Bloch body toward the uniform mixed-state steady state. 
    Other than $\gamma (C)$, the numerical parameters are the same as in Fig.~\ref{fig:pure_state_dynamics}.
    }
    \label{fig:mixed_state_dynamics}
\end{figure*}

Having established that homogeneous continuous monitoring with unit efficiency generates the uniform ensemble over pure states, we now turn to the mixed-state problem. In the state-independent setting, the purification tendency of the dynamics obstructs a uniform steady state throughout the Bloch body, for any fixed choice of $\gamma >0$ and $\eta<1$\footnote{See App.~\ref{app:state-independent_failure}.}. Any form of unitary evolution from Hamiltonian terms, even from state-dependent Hamiltonians, cannot counteract this purification either. A natural extension is therefore to allow additional decoherence channels to depend on the state.

\subsection*{State-dependent rates}

Motivated by the success of homogeneous monitoring in the generation of uniform pure-state ensembles,
we retain our attention on homogeneous control variables at the outset, but now allow non-zero decoherence rates that vary with the purity coordinate $C$ according to
\begin{align}\label{eq:Gamma_ansatz}
    \gamma_j=\gamma(C),
    \qquad
    l(C)=2k+2\gamma(C).
\end{align}
Since $\gamma(C)$ multiplies a Lindblad dissipator, physical consistency requires
\begin{align}
    \gamma(C)\geq 0
    \qquad
    \forall~C.
    \label{eq:Gamma_positive}
\end{align}
Homogeneous scalar rates depending only on the radial purity coordinate $C$ reflect the unitarily invariant nature of the state-space geometry and provide one of the simplest analytically tractable extensions beyond state-independent dynamics. More general scalar dependencies $\gamma(\rho)$ could also be considered, but are left for future investigation.

Allowing the decoherence rate to depend on the state is, at present, best viewed as a phenomenological extension, though one that is not without physical motivation. 

In continuously monitored systems, feedback based on real-time state estimation from the measurement record induces an effective state-dependent contribution to the dynamics, since the control applied at each time depends on the conditioned state  \cite{Andersson2022}.
For instance, in superconducting qubit platforms, Hamiltonian controls can be modulated in real time based on observed signals \cite{Vijay2012,Murch2016,RisteFeedback2012}. If, as is often considered in the literature \cite{Wiseman1994,Barchielli_2012}, Hamiltonian driving can depend on the state via non-Markovian state-estimation-based feedback, in theory a corresponding pure-dephasing dissipator can also be generated. Such a dissipator can be formed from a zero-mean noisy field applied to the system. If controls experience amplitude noise, then two equal-magnitude and opposite-sign control fields give rise to a pure dephasing dissipator with rate proportional to the control amplitude-squared.

Although such mechanisms do not straightforwardly realise a strictly state-dependent decoherence rate at the microscopic level, they suggest that, at the level of coarse-grained or feedback-engineered stochastic dynamics, behaviours analogous to the present model may be achievable.
Additionally, it is quite possible that, as an alternative to state-dependent dephasing rates, similar results could be achieved through the use of state-dependent efficiencies, $\eta (C)$. This option may be easier to implement in a lab setting.

\subsection*{Geometry of the mixed-state boundary}

In general, the boundary of the quantum state space is not fully characterised by the purity coordinate $C$. While $C=0$ identifies the pure-state manifold, for $d>2$ the boundary of the mixed-state body consists of all rank-deficient density matrices \cite{BengtssonZyczkowski}.

A representation-independent characterisation of the Bloch boundary can be created from the positivity of the state:
\begin{align} \label{eq:q_def}
    C_\phi(\rho):=\langle \phi|\rho|\phi\rangle \ge 0
    \qquad \forall\, |\phi\rangle\in\mathcal H.
\end{align}
The boundary is defined by the saturation of one or more of these constraints, i.e. by the existence of $|\phi\rangle\neq 0$ such that $C_\phi(\rho)=0$, or equivalently $|\phi\rangle\in\ker\rho$. For $d=2$, this reduces to the single condition $C=0$, corresponding to the Bloch sphere. For $d>2$, however, the boundary has a more intricate structure which cannot be captured by $C$ alone.

\subsection*{Conditions for mixed-state uniformity}

Consider the FPE \eqref{eq:FPE_covariant} with a boundary imposed by saturation of the inequality \eqref{eq:q_def}. A sufficient set of conditions for the uniform mixed-state distribution to be stationary and uniquely generated is:

\begin{enumerate}
    \item[\textbf{(I)}] \textbf{Stationarity of the uniform bulk density:}
    \begin{align}\label{eq:I}
        -\nabla_\nu h^\nu
        +\tfrac{1}{2}\nabla_\mu\nabla_\nu D^{\mu\nu}
        =0.
        \tag{I}
    \end{align}

    \item[\textbf{(II)}] \textbf{Invariance of the state space:}
    \begin{align}\label{eq:II}
        dC_\phi(\rho_t) \ge 0
        \tag{II}
    \end{align}
    whenever $C_\phi(\rho_t)=0$, for all $|\phi\rangle\in\mathcal H$.

    This necessitates no outward flux on the pure-state boundary:
    \begin{align}\label{eq:II*}
        n_\nu J^\nu = 0
        \qquad
        \text{on } C=0
        \tag{II$^*$}
    \end{align}
     where $n_\nu=\nabla_\nu C$.

    \item[\textbf{(III)}] \textbf{Non-degenerate diffusion in the interior:}
    \begin{align}\label{eq:III}
        v_\mu D^{\mu\nu}(x)v_\nu > 0
        \tag{III}
    \end{align}
    for all nonzero covectors $v_\mu$.
\end{enumerate}

Condition~\eqref{eq:I} ensures bulk stationarity. Condition~\eqref{eq:II} enforces invariance of the quantum state space, guaranteeing that the stochastic dynamics preserves positivity and trace---that valid density matrices map to valid density matrices. Condition~\eqref{eq:II*} is a geometric specialisation of \eqref{eq:II} to the pure-state boundary and is necessary but not sufficient in general dimensions. Condition~\eqref{eq:III} generalises the pure-state condition \eqref{eq:cond_iv} to mixed states, ensuring ellipticity of the diffusion and hence irreducibility of the dynamics in the interior.

\begin{figure}[t]
    \centering
    \includegraphics[width=0.95\columnwidth]{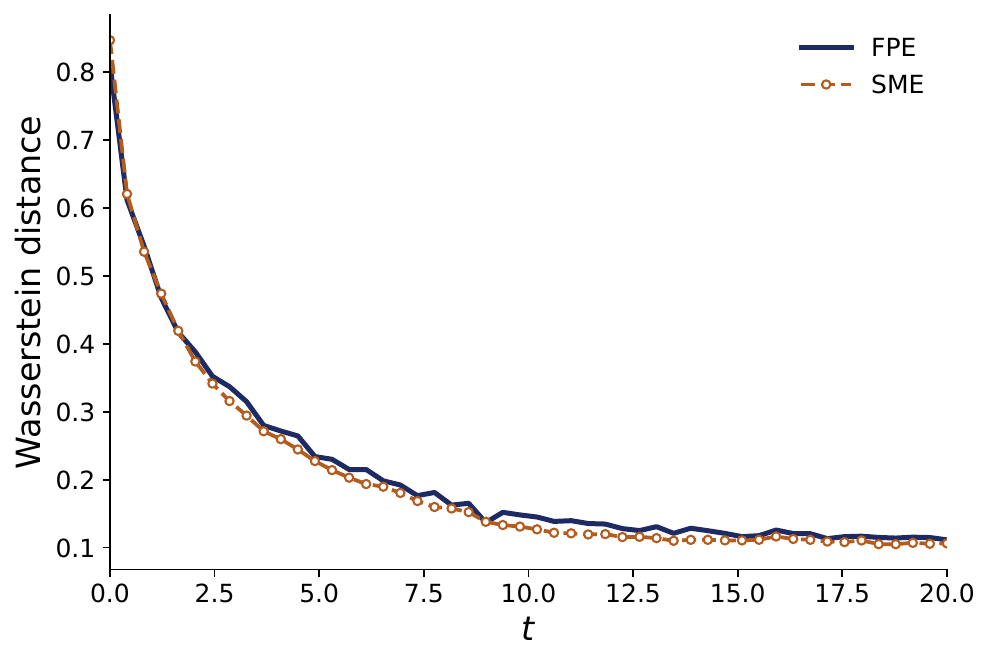}
    \caption{
    Wasserstein-1 distance (with entropic regularisation) to the uniform bulk reference distribution in the Bloch ball for the mixed-state protocol with $\gamma(C)=-6kC$. 
    The solid curve shows the Fokker--Planck result, while the dashed curve with markers shows the corresponding SME trajectory estimate. 
    The two curves show the same qualitative relaxation toward the uniform mixed-state distribution, with the SME result exhibiting the expected additional fluctuations associated with finite sampling.
    }
    \label{fig:mixed_state_wasserstein}
\end{figure}



\section{State-dependent rates}
\label{sec:MixedStateProofs}

In this section we verify conditions~\eqref{eq:I}, \eqref{eq:II}, \eqref{eq:II*}, and \eqref{eq:III} for the homogeneous purity-dependent ansatz introduced in Sec.~\ref{sec:MixedStates}, under the effectively Euclidean HS metric. Afterwards, we briefly study the same conditions under the Bures metric.

In this setting, the homogeneous drift becomes
\begin{align}
    f^\nu
    =
    \sum_{j,k} h_j f_{\nu jk}x_k
    -2d\,l(C)\,x_\nu,
    \label{eq:mixed_drift}
\end{align}
with partial derivative
\begin{align}
    \partial_\nu f^\nu
    =
    2d(1-d^2)\,l(C)
    -2d\,l'(C)\sum_j x_j^2.
    \label{eq:mixed_divf}
\end{align}
As provided in App.~\ref{app:FPE_terms}, the diffusion tensor remains unchanged from the homogeneous monitored case. Using the expression for its double partial derivative,
\begin{align}
    \tfrac{1}{2}\partial_\nu\partial_\mu D^{\mu\nu}
    =
    4k\bigg[
        (d^2 + 2)&(d^2+1)\sum_j x_j^2 \nonumber \\
        &
        -\frac{(3d^2+4)(d^2-1)}{d}
    \bigg],
    \label{eq:mixed_divdivD}
\end{align}
together with
\begin{align}
    \sum_j x_j^2 = \frac{2(d-1)}{d}+2C ,
    \label{eq:mixed_state_identity}
\end{align}
we proceed to demonstrate conditions \eqref{eq:I}--\eqref{eq:III}.

\subsection*{(I) Stationarity}

To begin, we impose condition~\eqref{eq:I} to find constraints on $\gamma (C)$. Substituting Eqs.~\eqref{eq:mixed_divf}, \eqref{eq:mixed_divdivD}, and \eqref{eq:mixed_state_identity} into \eqref{eq:I}, and using $l(C)=2k+2\gamma(C)$, gives a first-order linear inhomogeneous differential equation in $\gamma (C)$:
\begin{align}
    0
    =&
    -2(dC+d-1)\gamma'(C)
    +(d-d^3)\gamma(C)
    \nonumber\\
    &\hspace{0em}
    -2k\left[ d^4 - 2d^3 + 3d^2 - 4d + 2 + C(d^4 + 3d^2 + 2)\right].
    \label{eq:Gamma_ODE}
\end{align}
This can be simplified slightly by introducing
\begin{align}
    M:=dC+d-1,
\end{align}
so that $0\leq M\leq d-1$ over the entire interval $\frac{1-d}{d}\leq C\leq 0$. With this, Eq.~\eqref{eq:Gamma_ODE} takes the normal form
\begin{align}
    \gamma'(C)
    +\frac{d(d^2-1)}{2M}\gamma(C)
    =
    & \frac{-k}{Md}\big[
    (d^4 + 3d^2 + 2)M \nonumber\\ 
    & \ \ \  -
    (d^4 - d^3 + 3d - 2)
    \big].
    \label{eq:Gamma_normal_form}
\end{align}

This ODE can be solved using the integrating factor $\mu(C)=M^{(d^2-1)/2}$.
Multiplying Eq.~\eqref{eq:Gamma_normal_form} by $\mu(C)$ and integrating gives the general solution
\begin{align}
    \gamma(C)
    =
    a_\gamma \,(dC +d-1&)^\frac{-(d^2-1)}{2} \nonumber \\
    &+\frac{2k}{d^2}(d^2+2)\big(2-d-dC\big),
    \label{eq:Gamma_solution}
\end{align}
where $a_\gamma $ is an integration constant.

For $d>2$, this family exhibits a trade-off between regularity and positivity. The regular solution, $\gamma(C)\propto(2-d-dC)$, is negative at $C=0$, while any choice $a_\gamma\neq0$ introduces a divergence at the maximally mixed state, $C=(1-d)/d$. Thus, within the present homogeneous purity-dependent ansatz, higher-dimensional mixed-state ensembles cannot be realised by physically admissible decoherence rates. However, whether more general ansatzes, for example of the isotropic form $\gamma(C,\Tr\rho^3,\ldots,\Tr\rho^d)$, can overcome this obstruction remains an open question.

For $d=2$, Eq.~\eqref{eq:Gamma_solution} reduces to
\begin{align}
    \gamma(C)
    =
    a_\gamma (2C+1)^{-3/2}-6kC.
\end{align}
In particular, the regular solution $a_\gamma=0$ is
\begin{align}\label{eq:Gamma_solution_qubit}
    \gamma(C)=-6kC,
\end{align}
which is non-negative throughout the full interval $-\tfrac{1}{2}\leq C\leq 0$. Substituting Eq.~\eqref{eq:Gamma_solution_qubit} back into \eqref{eq:I} confirms that condition~\eqref{eq:I} is satisfied. Thus, the HS bulk-uniform mixed-state problem admits a regular and physically admissible qubit solution within the present purity-dependent homogeneous ansatz.


\subsection*{(II) Invariance of the state space}

If $C_\phi(\rho)=0$ for some $|\phi\rangle$, positivity requires that the evolution does not drive $C_\phi$ negative. Thus, on the boundary, it is sufficient to ensure that the combined stochastic and deterministic evolution satisfies $dC_\phi \ge 0$. This guarantees that the process cannot leave the admissible region $C_\phi\ge 0$, and hence preserves positivity. We now verify this condition.

First, we recall the SME as in \eqref{eq:SME_ABform}, and allow both the deterministic $A$ and stochastic $B$ generators to depend on $\rho$. In terms of $q$,
\begin{align}
    dC_\phi
    =
    \langle \phi|A(\rho)|\phi\rangle\,dt
    +
    \sum_j \langle \phi|B_j(\rho)|\phi\rangle\,dW_j.
\end{align}
If $|\phi\rangle\in\ker \rho$, we can use $\rho|\phi\rangle=0$ and $\langle\phi|\rho=0$ to find
\begin{align}
    \langle \phi|B_j(\rho)|\phi\rangle = 0
    \qquad \forall j,
\end{align}
where we have used the form of the superoperator $\mathcal H[S_j]\rho$ in $B_j (\rho)$.
So, the stochastic term has no component normal to the boundary.

For the drift term, the Hamiltonian contribution vanishes, while the dissipative term yields
\begin{align}
    \langle \phi|A(\rho)|\phi\rangle
    =&
    \sum_j l_j(\rho)\,\langle \phi|S_j\rho S_j|\phi\rangle \nonumber \\
    = &
    \sum_j l_j(\rho)\,\langle \psi_j|\rho|\psi_j\rangle \ge 0,
\end{align}
where $|\psi_j\rangle:=S_j|\phi\rangle$ and positivity of $\rho$ has been used. Hence, provided $l_j(\rho)\ge 0$, the deterministic evolution cannot drive $C_\phi$ negative. Furthermore, the evolution preserves trace exactly. Taking the trace of Eq.~\eqref{eq:SME_main}, we obtain
$\Tr(d\rho)=0 $,
so that $\Tr(\rho_t)=1$ for all times.

Thus, for all boundary states and all $|\phi\rangle\in\ker\rho$, the stochastic contribution vanishes and the deterministic drift is non-negative. It follows that the positivity constraints are preserved, and the dynamics leaves the convex set of density matrices invariant.

This condition replaces the geometric boundary-flux condition for $d>2$ and provides a complete demonstration of state space invariance. We emphasise that, due to the state dependence of $l(C)$, the resulting evolution is nonlinear in $\rho$ and therefore does not define a linear completely positive trace-preserving (CPTP) quantum channel in the usual sense. Nevertheless, the dynamics remains a well-defined effective stochastic state evolution because it preserves Hermiticity, trace, and positivity of the density operator.

\subsection*{(II$^*$) Pure-state boundary}

For a qubit, the Bloch boundary $C_\phi (\rho) = 0$ is given entirely by $C=0$, and condition~\eqref{eq:II} reduces to the geometric no-flux condition \eqref{eq:II*}. 

For a uniform density, the normal current is
\begin{align}
    n_\nu J^\nu
    =
    \varrho_*
    \left(
        f^\nu \partial_\nu C
        -
        \tfrac{1}{2}(\partial_\mu D^{\mu\nu})\partial_\nu C
    \right).
\end{align}
On the manifold $C=0$, requiring that the dynamics reduces to the homogeneous pure-state case fixes $\gamma(0)=0$. For the qubit solution $\gamma(C)=-6kC$, this holds automatically, and hence $n_\nu J^\nu = 0$ on $C=0$. Thus \eqref{eq:II*} is satisfied.

For $d>2$, imposing $\gamma(0)=0$ fixes the integration constant
\begin{align}
    a_\gamma  = -(d-1)^{\frac{d^2-1}{2}}\,
    \frac{2k}{d^2}(d^2+2)(2-d),
\end{align}
ensuring consistency with the pure-state boundary. This condition is necessary but not sufficient for full invariance, which is instead guaranteed by \eqref{eq:II}.

However, for $a_\gamma \neq 0$, the solution $\gamma(C)$ diverges at $C=\frac{1-d}{d}$, corresponding to the maximally mixed state. This singular behaviour raises additional regularity concerns, and suggests that the present construction yields a well-behaved uniform mixed-state ensemble most naturally in the qubit case. More generally, this obstruction is not entirely unexpected since, for $d>2$, the geometry of the mixed-state space is not fully characterised by the purity coordinate $C$ alone. Higher-order unitary invariants, such as $\Tr(\rho^3),\dots,\Tr(\rho^d)$, are required to distinguish inequivalent spectral structures. Restricting the homogeneous state-dependent ansatz to dependence only on $C$ therefore imposes a strong reduction of the available geometric degrees of freedom, which appears insufficient to realise regular uniform mixed-state ensembles in higher dimensions.

\subsection*{(III) Interior ellipticity of the diffusion tensor}

We now establish condition~\eqref{eq:III}, namely that the diffusion tensor is strictly positive definite in the interior of the Bloch body, that is,
for all mixed states with a full-rank density matrix.

As in the demonstration of \eqref{eq:cond_iv}, using the definition $D^{\mu\nu}=\sum_j \sigma^\mu_j \sigma^\nu_j$ from Eq.~\eqref{eq:diffusion_tensor_def}, we have
\begin{align}
    v_\mu D^{\mu\nu} v_\nu
    =
    \sum_j (v_\mu \sigma^\mu_j)^2 \geq 0.
    \label{eq:III_sum_squares}
\end{align}
Again, by construction, the diffusion is positive semidefinite. To prove strict positivity, it suffices to show that the kernel is trivial. That is, we must show that
\begin{align}
    v_\mu \sigma^\mu_j = 0
    \quad \forall j
    \qquad  \Longleftrightarrow \qquad v_\mu = 0.
    \label{eq:III_kernel_condition}
\end{align}

Substituting the homogeneous noise amplitudes in Eq.~\eqref{eq:noise_amplitudes_main}, this condition becomes
\begin{align}
    \tfrac{2}{d} v_j + g_{\mu j n} x_n v_\mu - (x \cdot v)\, x_j = 0
    \qquad \forall j,
    \label{eq:III_component_condition}
\end{align}
where $x \cdot v := x^\mu v_\mu$.

Rather than attempting to solve Eq.~\eqref{eq:III_component_condition} directly in Bloch coordinates, it is convenient to recast it in operator form. Define the traceless Hermitian operator $V := \tfrac{1}{2} v_\mu S_\mu$. Using the Bloch decomposition \eqref{eq:bloch_rep} and the $SU(d)$ anticommutation relations \eqref{eq:SUd_algebra}, one finds that Eq.~\eqref{eq:III_component_condition} is equivalent to\footnote{See App.~\ref{app:III_derivation}.}
\begin{align}
    \{\rho, V\} = 2 \Tr(\rho V)\, \rho.
    \label{eq:III_operator_condition}
\end{align}

We now assume that $\rho$ lies in the interior of the Bloch body, so that it is full rank. Diagonalising $\rho$ and expressing Eq.~\eqref{eq:III_operator_condition} in its eigenbasis yields
\begin{align}
    (\lambda_i + \lambda_j) V_{ij}
    =
    2 \Tr(\rho V)\, \lambda_i \delta_{ij},
\end{align}
where $\lambda_i > 0$ are the eigenvalues of $\rho$. Since $\lambda_i + \lambda_j > 0$ for all $i,j$, it follows that $V_{ij}=0$ for $i\neq j$, while the diagonal elements satisfy $V_{ii}=\Tr(\rho V)$ for all $i$. Hence $V \propto \mathds{1}$, and since $V$ is traceless, we conclude that $V=0$. Therefore $v_\mu=0$.

This proves that the kernel of $D^{\mu\nu}$ is trivial in the interior, and hence that the diffusion tensor is strictly positive definite. Condition~\eqref{eq:III} therefore holds for all full-rank mixed states. We emphasise that this argument relies on the full-rank property of $\rho$, and therefore applies only in the interior of the Bloch body. At the boundary, where $\rho$ becomes rank-deficient, degeneracies of the diffusion tensor may occur. This is consistent with the formulation of condition~\eqref{eq:III}, which is imposed only in the interior, while boundary behaviour is controlled separately by condition~\eqref{eq:II}.

\medskip

This completes the verification of conditions~\eqref{eq:I}, \eqref{eq:II}, \eqref{eq:II*}, and \eqref{eq:III}. The dynamics, in principle, therefore admits the uniform mixed-state distribution as a stationary state, and—subject to the ellipticity and invariance conditions above—generates it dynamically from generic initial conditions. For a qubit, these conclusions are evidenced in Fig.~\ref{fig:mixed_state_dynamics} where, in Fig.~\ref{fig:mixed_state_wasserstein}, the distance between $\varrho$ and the HS uniform density monotonically decreases as the dynamics evolve. We reiterate, however, that for $d>2$, the purity-dependent solution of $\gamma (C)$ becomes unstable.

\subsection*{Bures-uniform mixed states}

We now briefly consider the mixed-state uniformity problem with respect to the Bures metric, restricting our attention to the implications of the stationarity condition \eqref{eq:I} for $d=2$ systems. Additional derivation steps can be found in App.~\ref{app:BuresDetails}.

Using a simplified homogeneous diffusion tensor for a qubit, one obtains
\begin{align}
    \nabla_\nu \nabla_\mu D^{\mu\nu}
    =
    32k(4r^2-3)
\end{align}
while the covariant divergence of the deterministic drift in Eq.~\eqref{eq:mixed_drift} is
\begin{align}
    \nabla_\nu f^\nu
    =
    l(C)\left[-12+\frac{2(1+2C)}{C}\right]
    -4(1+2C)l'(C).
\end{align}
Lastly, the divergence of the geometric drift is
\begin{align}
    \nabla_\nu \big( \tfrac12 \Gamma^\nu_{ij} D^{ij}\big) 
    &= 4k\big(3-8C-\tfrac1C \big).
\end{align}
Substituting into \eqref{eq:I} gives a first-order linear ODE for $\gamma_B(C)$,
\begin{align}\label{eq:ODE_BuresGamma}
    \gamma_B'(C)( 4C+2) = \gamma_B (C)\big(\tfrac1C -4 \big) - 5k(1+8C).
\end{align}
We seek a trial solution of the form $\gamma_B(C)=bC$.
Substituting this ansatz into Eq.~\eqref{eq:ODE_BuresGamma} gives
\begin{align}
    \gamma_B(C)
    = -5kC,
    \label{eq:Gamma_Bures_solution}
\end{align}
which is non-negative throughout the Bloch ball. Interestingly, Eq.~\eqref{eq:Gamma_Bures_solution} shares a very similar form to that of the HS solution in Eq.~\eqref{eq:Gamma_solution_qubit};
compared with the HS solution, the smaller $\gamma_B (C)$ rate reflects a stronger concentration of the Bures measure near the pure-state boundary.


\section{Post-mixing protocol}\label{sec:inefficient}

\begin{figure*}[t]

    \centering

    \includegraphics[width=\textwidth]{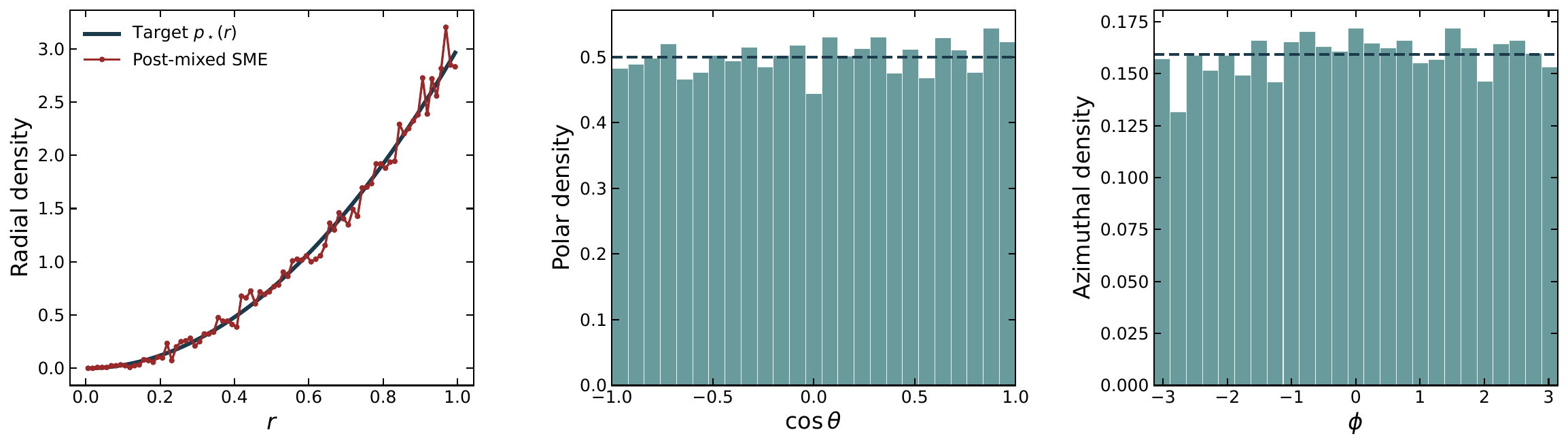}

    \vspace{0.0cm}

    \includegraphics[width=\textwidth]{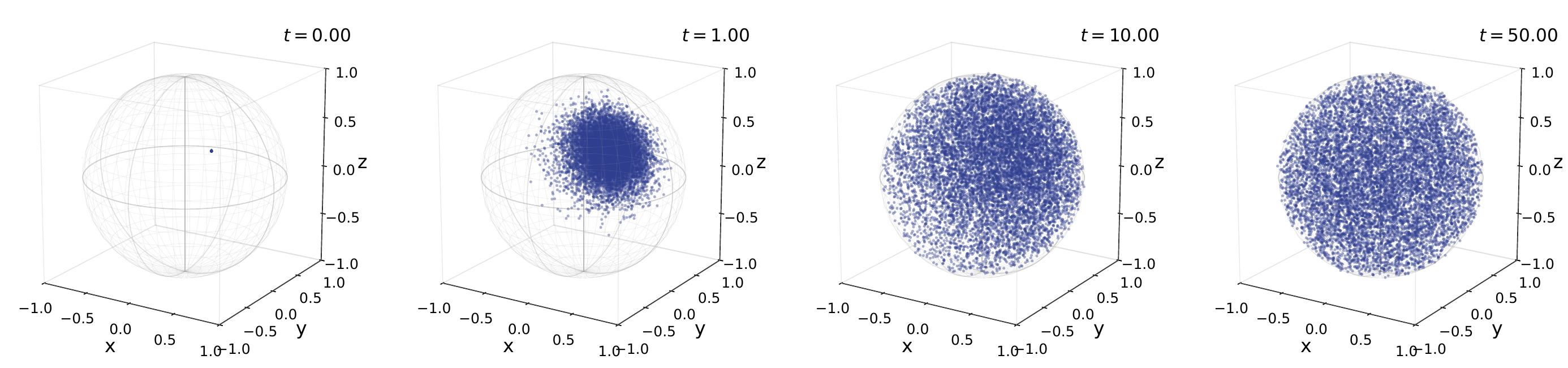}



    \caption{Post-mixed construction of the HS mixed-state ensemble. 
    The top row shows final-time SME diagnostics: the radial density compared with the HS target $p_{\mathrm{HS}}(r)=3r^2$, together with the polar and azimuthal angular distributions. The approximately flat angular histograms indicate that the post-mixed ensemble is isotropic, while the radial panel tests the desired volume measure. 
    The middle row shows time-resolved SME trajectory clouds for the post-mixed ensemble. 
    The calculation uses isotropic inefficient monitoring with $\gamma=0$, measurement strength $k=10^{-2}$, time step $dt=5\times10^{-3}$, final time $t_{\mathrm{final}}=50$, and initial Bloch vector $(x,y,z)=(0.3,0,0.3)$. In calculating the efficiency weights, $10,000$ trajectories were used, sorted into $81$ radial bins with regularisation parameter $\lambda = 10^{-8}$.
    }
    \label{fig:postmix_hs_dynamics}
\end{figure*}

\begin{figure}[t]
    \centering
    \begin{overpic}[width=0.95\columnwidth]{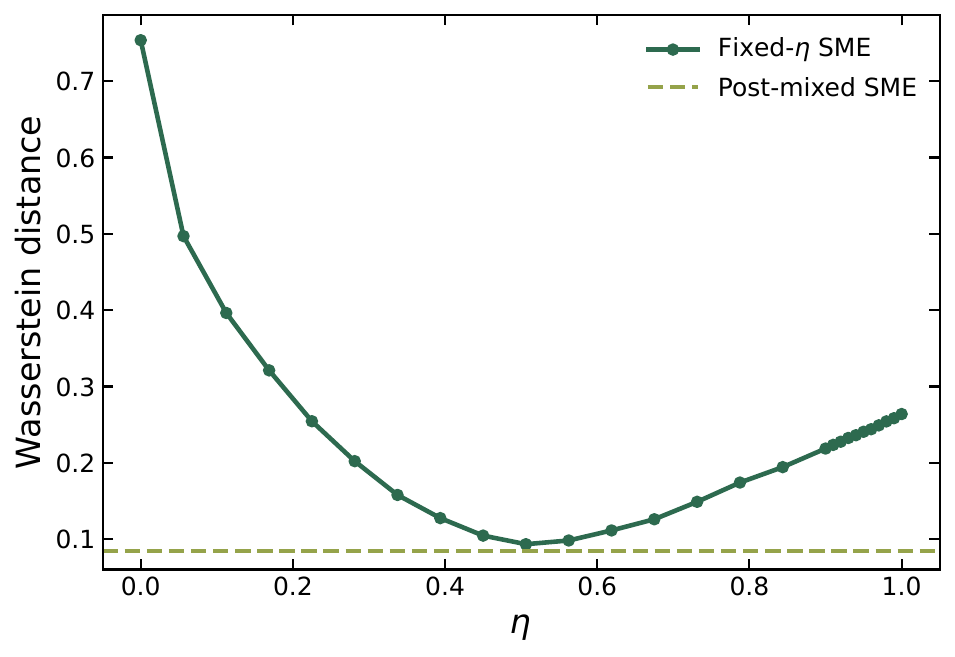}
    
        \put(37,22){\includegraphics[width=0.45\columnwidth]{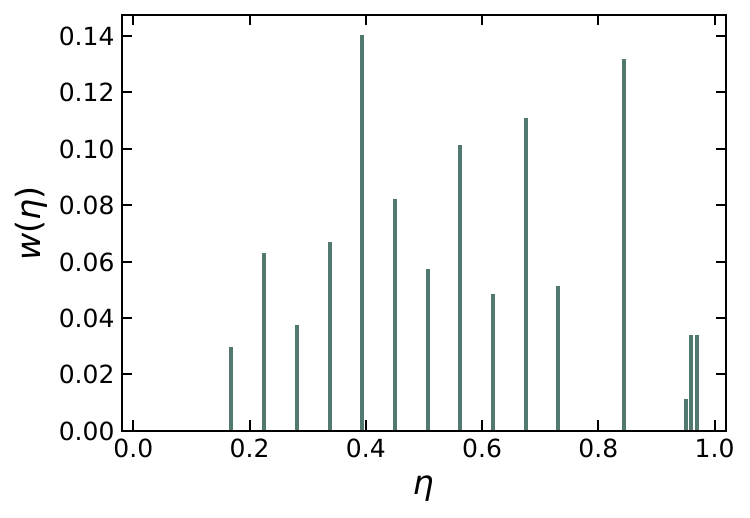}
        }
        
    \end{overpic}
    \caption{
    Final-time Wasserstein distance to the HS reference ensemble for fixed-efficiency SME ensembles, compared with the post-mixed SME ensemble. 
    The solid curve shows the Wasserstein distance obtained using a single fixed efficiency $\eta$, while the dashed horizontal line gives the corresponding post-mixed value obtained after optimising the sampling weights $w(\eta)$. 
    The inset shows the inferred efficiency weights used to construct the HS post-mixed ensemble. Note that $\eta<0.2$ receive a weighting of zero, suggesting that detailed-balance radii $r_\eta^2$ have more influence on state reconstruction than $r_L^2$. 
    More $\eta$ candidates were chosen between $0.9$--$1.0$ to reflect the precision required to reconstruct precise boundary behaviour. In calculating the Wasserstein distance, 4000 points were sampled from the desired uniform distribution and an equivalent number of post-mixing trajectories were used to calculate the Wasserstein distance. Other parameters are as in Fig.~\ref{fig:postmix_hs_dynamics}.
    }
    \label{fig:postmix_hs_wasserstein_weights}
\end{figure}

The construction developed above shows that a uniform mixed-state ensemble can be generated exactly, at least formally, by introducing a state-dependent decoherence rate $\gamma(C)$. This result is useful as a proof of principle: it demonstrates that the radial probability current (that is, evolution towards the boundary of the Bloch body) can be engineered so that the interior uniformity measure is dynamically selected. However, it is also the least experimentally appealing part of the proposal. Implementing $\gamma(C)$ requires real-time knowledge of the conditional purity coordinate $C$, together with sufficiently accurate and rapidly tunable dissipative channels. We therefore now ask whether a sampling of state-independent trajectories can reproduce the same target ensembles, whilst being more realistic in a lab setting.

In this section, we focus on the case that $d=2$, though we hope that the approach and discussion may be used to inform analysis of more complex systems.

\subsection*{Properties of $\eta <1$ distributions}

A natural candidate to sample from is measurement efficiency. Efficiency is experimentally familiar: it may arise from finite collection efficiency, amplifier noise, loss before detection, or deliberately discarding a fraction of the measurement record \cite{guryanova2020ideal}. In many experimental platforms, the effective measurement efficiency may also be partially tuned through the choice of filtering, amplification, and signal-processing protocols \cite{Lecocq2021,Eddins2018}.

We consider simultaneous monitoring along the three Pauli directions with a fixed efficiency $\eta<1$. We set $\gamma=0$ throughout this section and consider the isotropic inefficient SME
\begin{align}\label{eq:inefficieny_SME}
    d\rho
    =
    2k\sum_{j}\mathcal{D}[S_j]\rho\,dt
    +
    \sqrt{2\eta k}\sum_{j}\mathcal{H}[S_j]\rho\,dW_j 
\end{align}
such that the corresponding Bloch-vector dynamics remain rotationally symmetric, but the balance between measurement-induced purification and isotropic decoherence is shifted. 

We choose to express the associated FPE in Cartesian Bloch coordinates, governing the evolution of a probability density with respect to the Euclidean (HS) volume element. Accordingly, all differential operators appearing below are ordinary partial derivatives in these coordinates, and the Fokker-Planck drift vector again coincides with the deterministic drift, $h^\nu=f^\nu$. This should be distinguished from the covariant formulations used in Sec.~\ref{sec:MixedStateProofs}, where the dynamics are posed relative to a chosen metric throughout the evolution. Here, by contrast, the geometry of the target ensemble is not relevant to the dynamics itself, and will enter only later through the post-mixing construction outline below.

To investigate properties of the ensuing FPE, we first look at potential detailed balance conditions, by repeating a similar analysis used to demonstrate \eqref{eq:cond_i} and \eqref{eq:cond_iii_strong}. In this way\footnote{See App.~\ref{app:FPE_terms}.}, we find that for zero Hamiltonian,
\begin{align}
    f^\nu - \tfrac{1}{2}\partial_\mu D^{\mu \nu} = &-4 d k x_\nu+ 4 k\eta \bigg[
    \frac{3d^2+4}{d} x_\nu  \nonumber\\
    & \hspace{2cm} + (d^2+2) \sum_{j,m}g_{\nu jm}x_m  x_j \nonumber\\
   & \hspace{2cm}  - (d^2+2) x_j^2 x_\nu 
    \bigg].
\end{align}
In the qubit case, using $r^2 = \sum_j x_j^2$, the detailed-balance expression vanishes at the radius $r_\eta$ which obeys
\begin{align}
    r_\eta^2
    =
    \frac{4\eta-1}{3\eta},
    \qquad
    \eta\in\left[\tfrac{1}{4},1\right],
    \label{eq:reta}
\end{align}
so that $\eta=1/4$ corresponds to the maximally mixed state and $\eta=1$ to the pure-state boundary. So, for a single fixed value of $\eta$, the deterministic balance condition predicts a preferred radius, corresponding to a purity $C_\eta$ (in fact, since we are reviewing only qubit systems, we could frame this entire approach in terms of $C$ but choose to speak in term of $r$ to facilitate visualisation and to simplify metric representations used later). Thus, by varying $\eta$, the full HS radial interval $0\leq r\leq 1$ can in principle be spanned. This observation is central to the construction below.

However, Eq.~\eqref{eq:reta} should not be over-interpreted. It does not imply that each fixed-$\eta$ dynamics produces an infinitely sharp shell, since the steady state condition \eqref{eq:cond_iii} at $r_\eta$ does not hold. The inefficient dynamics obey a weaker set of radial properties than those used in the pure-state construction. It is useful to separate these properties, defined with respect to the Cartesian Bloch-coordinate dynamics, as follows.

\begin{enumerate}
    \item[(i$'$)]  \textbf{Local detailed balance:}
    \begin{align}\label{eq:i'}
        f^\nu - \tfrac{1}{2}\partial_\mu D^{\mu\nu}=0
        \tag{i$'$}
    \end{align}
    There exists a radius $r_\eta$ at which the radial probability current vanishes. Equivalently, the local balance condition is satisfied on the shell $C=C_\eta$.
    
    \item[(ii$'$)] \textbf{Isotropy:}
    \begin{align}\label{eq:ii'}
        v_\mu D^{\mu \nu}v_\nu >0 \quad \text{for all nonzero }v .
        \tag{ii$'$}
    \end{align}
    The drift and diffusion tensors remain invariant under global rotations of the Bloch vector. Hence any long-time distribution generated from generic initial data is angularly uniform, up to sampling fluctuations.

    \item[(iii$'$)] \textbf{Moment-balance radius:}
    \begin{align}\label{eq:iii'}
        &L(r^2-r_L^2)<0
        \quad \text{for }(r^2>r_L^2), \nonumber \\
        &L(r^2-r_L^2)=0
        \quad \text{for }(r^2=r_L^2),\text{ and}   \tag{iii$'$}\\
        &L(r^2-r_L^2)>0
        \quad \text{for }(r^2<r_L^2), \nonumber
    \end{align}
    for some radius $r_L \neq r_\eta$ in general. 
    \end{enumerate}
We have already demonstrated \eqref{eq:i'}, and \eqref{eq:ii'} follows from the validation of \eqref{eq:III}, since the diffusion tensor here differs only by a positive scalar factor $\eta$.

\begin{figure*}[t]
    \centering
    \includegraphics[width=\textwidth]{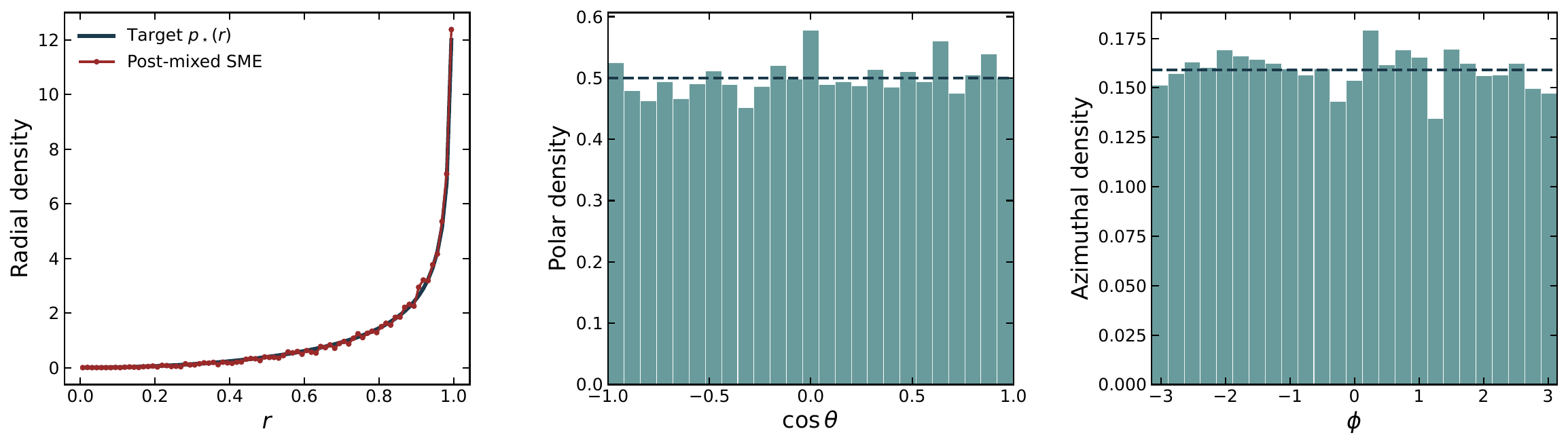}
    
    
    \caption{
    Post-mixed construction of the Bures mixed-state ensemble, showing the final-time radial and angular SME diagnostics. 
    The radial distribution is compared with the Bures target 
    $p_{\mathrm{B}}(r)=\frac{4}{\pi}r^2/\sqrt{1-r^2}$, while the polar and azimuthal panels test angular isotropy. 
    The enhanced boundary weight of the Bures measure is visible as a stronger concentration near large radii compared with the HS construction. 
    Parameters are as in Fig.~\ref{fig:postmix_hs_wasserstein_weights}.
    }
    \label{fig:postmix_bures_dynamics}
\end{figure*}

In reaching \eqref{eq:iii'} for qubits, an explicit evaluation of the backward generator yields
\begin{align}\label{eq:iii'_evaluated}
    \tfrac{1}{8k}L r^2
    =
    3\eta - 2(\eta+1)r^2 + \eta r^4 .
\end{align}
Thus \eqref{eq:iii'} is satisfied, with $r_L^2$ defined by the root at which the mean radial growth vanishes.

In general, however, this radius does not coincide with the detailed-balance radius $r_\eta$, reflecting the distinction between moment balance and probability-current balance in the presence of multiplicative diffusion. The two criteria probe different aspects of the dynamics: the former governs the mean evolution of $r^2$, while the latter determines where the radial probability current vanishes for a uniform density.

To make this distinction explicit, compare the radial probability current
\begin{align}
    x_\nu J^\nu
    =
    x_\nu f^\nu
    -
    \tfrac{1}{2}x_\nu \partial_\mu D^{\mu\nu},
\end{align}
with the generator applied to the squared radius,
\begin{align}
    Lr^2
    =
    2x_\nu f^\nu
    +
    D^\mu_{\mu},
\end{align}
where $\partial_\nu r^2=2x_\nu$ and $\partial_\mu\partial_\nu r^2=2\delta_{\mu\nu}$. These expressions coincide only under the additional structural relation
\begin{align}\label{eq:D_equivalence}
    D^\mu_{\mu}
    =
    -x_\nu \partial_\mu D^{\mu\nu}.
\end{align}
When this identity holds, the radii selected by current balance and moment balance are the same. This is the case for the unit-efficiency pure-state dynamics, where both conditions identify the boundary $r=1$. For inefficient measurements, however, Eq.~\eqref{eq:D_equivalence} is not generally satisfied at any specified radius, and the two notions of balance separate. 

The coexistence of \eqref{eq:i'} and \eqref{eq:iii'}, with mismatched $r_L $ and $ r_\eta$, implies that the dynamics does not collapse onto a single radius. Instead, trajectories experience competing effects: a vanishing probability current at $r_\eta$, and vanishing mean radial drift at $r_L$. The resulting steady-state behaviour is therefore a finite-width radial distribution rather than a delta function. We denote this conditional radial distribution in the following property.

\begin{enumerate}
    \item[(iv$'$)] \textbf{Finite radial kernel}\\
    \begin{align}
        P(r|\eta),
        \tag{iv$'$}
        \label{eq:iv'}
    \end{align}
    is the probability density of observing radius $r$ at the final readout time, given that the trajectory was evolved with efficiency $\eta$.
\end{enumerate}

The key conclusion from properties \eqref{eq:i'}-\eqref{eq:iv'} is that each $\eta $ results in an isotropic distribution, with a broad radial distribution governed by the efficiency by which it is measured.

\subsection*{Post-mixing}

The single-$\eta$ SME construction is not an exact mechanism for producing a target radial measure. Instead, different $\eta$ together provide a physically accessible family of isotropic radial kernels. A desired mixed-state ensemble can then be assembled by classical post-mixing over trajectories generated with different efficiencies.

Let $w(\eta)$ denote the probability density over efficiencies. The radial density of the post-mixed ensemble is then
\begin{align}
    p_{\mathrm{mix}}(r)
    =
    \int_{\eta_{\min}}^{\eta_{\max}}
    P(r|\eta)\, w(\eta)\, d\eta ,
    \label{eq:postmix_integral}
\end{align}
where, for the inefficient-measurement construction considered here, $\eta_{\min}=0$ and $\eta_{\max}=1$. The inverse problem is therefore to find $w(\eta)$ such that $p_{\mathrm{mix}}(r)\simeq p_*(r)$, for a given target radial density $p_*(r)$.

In practice we discretise Eq.~\eqref{eq:postmix_integral}. For a grid of efficiencies $\{\eta_a\}_{a=1}^{N_\eta}$ and radial bins $\{r_i\}_{i=1}^{N_r}$, the conditional kernel is estimated as
\begin{align}
    K_{ia}
    \simeq
    P(r_i|\eta_a),
\end{align}
from either SME trajectories or Fokker--Planck evolution. The sampling weights $\{w_a\}$ are then obtained by numerically solving the constrained least-squares problem \cite{lawson1995solving}
\begin{align}
    \min_{\{w_a\}}
    \sum_i
    \bigg(
        \sum_a K_{ia}w_a - p_*(r_i)
    \bigg)^2\Delta r
    +\lambda\sum_a w_a^2,
\end{align}
subject to $w_a\geq0$ and $\sum_a w_a=1$.
The regularisation parameter $\lambda$ suppresses spurious oscillations in the inferred weights without affecting the target ensemble in the large-sample limit.

For the HS measure on the Bloch ball, the volume element is\footnote{See App.~\ref{app:radial_densities}.}
$dV_{\mathrm{HS}}
    =
    4\pi r^2\,dr$,
which gives the radial density
\begin{align}
    p_{\mathrm{HS}}(r)=3r^2 .
    \label{eq:HS_radial}
\end{align}
For the Bures measure, the radial volume element is enhanced near the boundary \cite{BengtssonZyczkowski,HUBNER1992239},
\begin{align}
    p_{\mathrm{B}}(r)
    =
    \frac{4}{\pi}
    \frac{r^2}{\sqrt{1-r^2}} .
    \label{eq:Bures_radial}
\end{align}
Unlike the state-dependent $\gamma(C)$ construction, the divergence in Eq.~\eqref{eq:Bures_radial} does not require a divergent physical rate; it is reproduced through the weighting $w(\eta)$, which concentrates probability near $\eta\simeq1$. 

\begin{figure}[t]
    \centering
    \begin{overpic}[width=0.95\columnwidth]{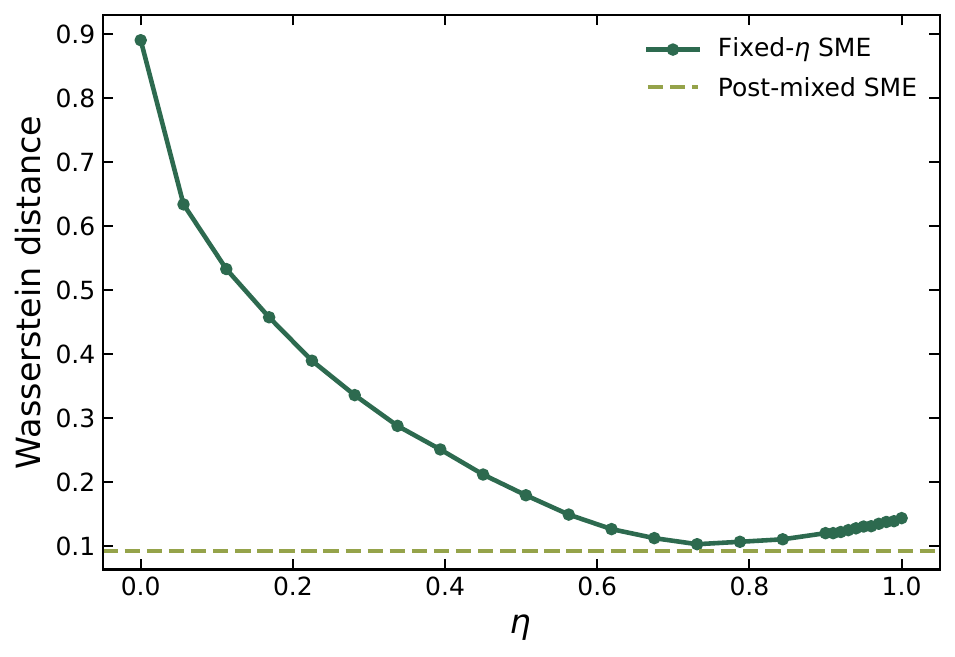}

        \put(48,21){\includegraphics[width=0.46\columnwidth]{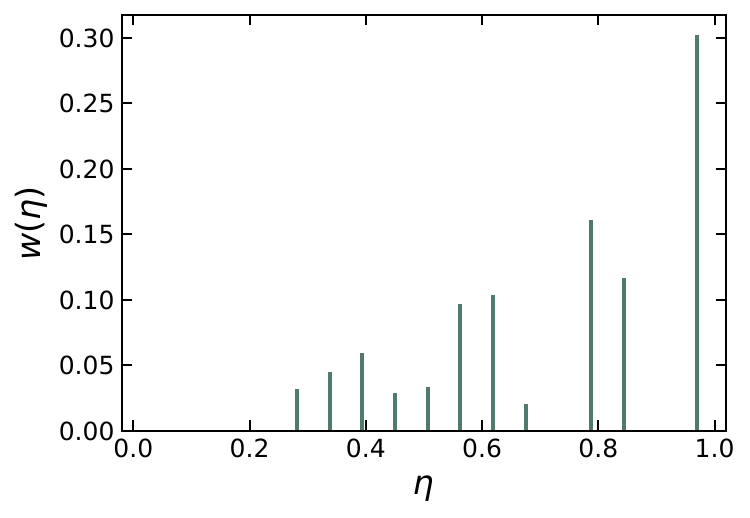}
        }
    \end{overpic}
    \caption{
    Final-time Wasserstein distance to the Bures reference ensemble for fixed-efficiency SME ensembles, compared with the post-mixed SME ensemble. The solid curve shows the fixed-$\eta$ result, while the dashed horizontal line shows the post-mixed value. The inset gives the inferred efficiency weights for the Bures target. 
    Compared with the HS case, the Bures radial density is more strongly weighted towards the pure-state boundary, so the optimisation correspondingly favours larger efficiencies. 
    The same SME parameters and initial state are used as in Fig.~\ref{fig:postmix_hs_wasserstein_weights}.
    }
    \label{fig:postmix_bures_wasserstein_weights}
\end{figure}

The resulting constructions are shown in Figs.~\ref{fig:postmix_hs_dynamics} and \ref{fig:postmix_hs_wasserstein_weights} for the HS case and Figs.~\ref{fig:postmix_bures_dynamics} and \ref{fig:postmix_bures_wasserstein_weights} for the Bures case. 
Figures~\ref{fig:postmix_hs_dynamics} and \ref{fig:postmix_bures_dynamics} show the corresponding ensemble structure. The top panels confirm that the post-mixed radial distributions closely reproduce the target densities in Eqs.~\eqref{eq:HS_radial} and \eqref{eq:Bures_radial}, while the angular distributions remain approximately uniform, demonstrating isotropy. The SME trajectory snapshots illustrate how the ensemble spreads through the Bloch ball.
The stronger boundary weighting in the Bures case is visible in both representations.

In Figs.~\ref{fig:postmix_hs_wasserstein_weights} and \ref{fig:postmix_bures_wasserstein_weights}, the fixed-$\eta$ Wasserstein distance exhibits a nontrivial dependence on efficiency, while the post-mixed ensemble achieves a strictly lower value, indicating improved agreement with the target measure. The insets show the inferred weights $w_a$, which are broadly distributed for the HS target but become increasingly concentrated towards $\eta\simeq1$ for the Bures case.

This post-mixing method should be interpreted differently from the exact $\gamma(C)$ protocol. It does not define a single autonomous dynamics with the desired stationary state. Instead, it constructs the target ensemble as a convex mixture of experimentally accessible processes. This replaces state-dependent dissipative feedback with classical randomisation over a single scalar parameter $\eta$, and the only object that must be characterised is the kernel $P(r|\eta)$.

The distinction between pre-mixing and post-mixing is important. In a pre-mixed protocol, one would vary $\eta$ during the evolution of a single trajectory. This defines a new SME with nontrivial time-dependent dynamics and does not, in general, produce a convex mixture of fixed-$\eta$ radial kernels. By contrast, in post-mixing one samples a value of $\eta$ at the beginning of each experimental run, evolves the trajectory with that fixed efficiency, and only then combines the resulting ensemble data. Operationally, one draws $\eta$ from a prescribed distribution, repeats the experiment many times, and pools the final states.

The same logic extends beyond efficiency sampling. Since increasing fixed decoherence $\gamma>0$ also shifts the effective radial distribution inward, one may similarly define a kernel $P(r|\gamma)$ and solve for a post-mixing distribution over $\gamma$\footnote{See App.~\ref{app:Gamma_postmixing}.}. More generally, any experimentally tunable parameter that generates an isotropic family of radial kernels that spans the entire state space may be used in Eq.~\eqref{eq:postmix_integral}.


\section{Discussion}
\label{sec:Discussion}

In this work, we have shown that homogeneous continuous monitoring provides a simple and physically natural mechanism for generating and preserving uniform ensembles of pure quantum states. By deriving the Bloch-space Langevin and Fokker--Planck equations associated with a general stochastic master equation, and analysing their geometric structure, we identified conditions sufficient for the emergence of a uniform steady state on the pure-state manifold. These conditions are satisfied for homogeneous monitoring with equal measurement strengths and unit efficiency.
Operationally, this provides a direct mechanism for generating $d$-dimensional Haar-random pure states through continuous monitoring alone, without the need to sample random unitaries.

More broadly, our analysis reveals a close connection between stochastic quantum dynamics and geometric diffusion. The purity coordinate $C=\Tr(\rho^2)-1$ plays the role of a Lyapunov-like function: under the backward Fokker-Planck generator $L$, one finds $LC>0$ for all mixed states, implying a systematic drift toward the pure-state boundary. This reflects the information gain associated with continuous measurement, which induces radial purification while diffusion explores the pure-state manifold.

The mixed-state problem is more subtle. In the state-independent setting, the same purification mechanism obstructs the existence of a uniform steady state throughout the Bloch body. Allowing the decoherence rate to depend on the purity coordinate $C$ modifies this structure. In the HS geometry, the requirement of bulk stationarity reduces to a first-order differential equation for $\gamma(C)$, which admits the regular, non-negative solution $\gamma(C)=-6kC$ for a qubit. A similar solution $\gamma(C)=-5kC$ generates a Bures-uniform ensemble of qubit states.
Thus, for qubits, this establishes the existence and uniqueness of the uniform mixed-state steady state within the present framework.

However, for $d>2$, while formal stationarity solutions for $\gamma(C)$ exist, the chosen ansatz exhibits a physical obstruction: admissible solutions for $\gamma(C)$ either become negative or develop singularities. Nevertheless, these obstructions arise within the restricted homogeneous purity-dependent ansatz considered here, and it remains possible that more general state-dependent constructions $\gamma(\rho)$ could support generation of higher-dimensional uniform mixed-state ensembles.


Rather than engineering a single autonomous dynamics, we introduced an alternative post-mixing construction. This approach constructs the target ensemble as a convex mixture of experimentally accessible stochastic processes with different measurement efficiencies.
The desired radial distribution is recovered by numerically characterising the conditional kernels and solving a constrained inverse problem for the sampling weights. We showed that, for qubits, both HS and Bures radial measures can be accurately reproduced.

Simulations of the post-mixed ensembles exhibit isotropic angular distributions and radial densities in close agreement with the target measures, as well as systematically reduced Wasserstein distance compared to any fixed-efficiency ensemble. In the HS case, the inferred weights are broadly distributed across efficiencies, whereas in the Bures case they concentrate near $\eta\simeq1$, reflecting the enhanced boundary weighting of the target measure. Importantly, this construction avoids the need for state-dependent decoherence, replacing it with classical sampling of a single experimentally tunable parameter. While weaker in a dynamically-exact sense, post-mixing may be more experimentally realistic and flexible, requiring only knowledge of the kernel $P(r|\eta)$.


Several directions for future work arise. Foremost is the extension of the mixed-state construction beyond qubits. This may require more general monitoring schemes, including anisotropic or state-dependent rates $\gamma(\rho)$, measurement strengths $k(\rho)$, and efficiencies $\eta(\rho)$, which could also enable a broader class of target ensembles. Alternatively, it would be valuable to establish, both analytically and numerically, whether the post-mixing construction can be generalised to arbitrary dimensions. Beyond these questions, a more complete and physically grounded picture may emerge from hybrid classical-quantum descriptions \cite{layton2025restoringsecondlawclassicalquantum}, which could clarify the physical plausibility of state-dependent decoherence. Finally, approximately uniform ensembles might instead be generated using Markovian feedback control, offering a potentially more experimentally accessible route at the expense of incomplete information about $\rho$.

Overall, these results demonstrate that continuous measurement is not only a tool for state estimation, but also a powerful mechanism for engineering statistical ensembles of quantum states. For pure states, it provides a complete and robust route to Haar-uniform ensembles. For mixed states, it reveals a richer landscape: where, at least for qubits, both purity-dependent decoherence and post-mixing constructions can be used to generate random states according to a chosen geometry.

\medskip
\acknowledgements

This work was supported by the Engineering and Physical Sciences Research Council [grant number EP/W524347/1]. HJDM acknowledges funding from a Royal Society Research Fellowship (URF/R1/231394). 
The codebase that supports the findings of this article is openly available \cite{data_QFPE}.

\bibliography{apssamp}


\clearpage
\onecolumngrid
\appendix

\section{Weak measurements for non-commuting observables}
\label{app:weakMeasurementCompatibility}

In this appendix we briefly justify the simultaneous use of weak measurements of non-commuting observables, in contrast to the incompatibility of sharp projective measurements.

\subsection{Projective measurements}

Consider two observables with spectral decompositions
\begin{align}
    A=\sum_a aP_a,
    \qquad
    B=\sum_b bQ_b,
\end{align}
where $\{P_a\}$ and $\{Q_b\}$ are orthogonal projectors. The associated nonselective measurement channels are
\begin{align}
    \Phi_A(\rho)=\sum_a P_a \rho P_a,
    \qquad
    \Phi_B(\rho)=\sum_b Q_b \rho Q_b.
\end{align}

Composing these maps yields
\begin{align}
    \Phi_A \circ \Phi_B(\rho)
    &= \sum_{a,b} P_a Q_b\, \rho\, Q_b P_a, \\
    \Phi_B \circ \Phi_A(\rho)
    &= \sum_{a,b} Q_b P_a\, \rho\, P_a Q_b.
\end{align}
These expressions coincide for all $\rho$ if and only if
\begin{align}
    [P_a, Q_b]=0
    \qquad \forall a,b,
\end{align}
i.e. if and only if the observables commute. Thus, sharp measurements of non-commuting observables are incompatible at the level of their induced quantum channels.

\subsection{Weak measurements}

By contrast, weak measurements are described by Kraus operators close to the identity. For a Hermitian observable $A$, a symmetric weak measurement may therefore be written as
\begin{align}
    K^{(A)}_\pm
    =
    \frac{1}{\sqrt{2}}
    \left(
        \mathds{1}
        \pm \varepsilon A
        - \frac{\varepsilon^2}{2}A^2
    \right)
    + O(\varepsilon^3),
\end{align}
where $\varepsilon \ll 1$ controls the measurement strength.

The corresponding nonselective channel is
\begin{align}
    \mathcal{E}_A(\rho)
    &=
    \sum_\pm K^{(A)}_\pm \rho K^{(A)\dagger}_\pm \\
    &=
    \rho
    + \varepsilon^2 \mathcal{D}[A]\rho
    + O(\varepsilon^3),
\end{align}
where $\mathcal{D}[A]\rho = A\rho A - \tfrac{1}{2}\{A^2,\rho\}$ is the Lindblad dissipator. Similarly, for an observable $B$, $\mathcal{E}_B(\rho) =
    \rho
    + \varepsilon^2 \mathcal{D}[B]\rho
    + O(\varepsilon^3)$.

We now compare the compositions. Expanding to leading nontrivial order,
\begin{align}
    \mathcal{E}_B \circ \mathcal{E}_A(\rho)
    &=
    \rho
    + \varepsilon^2\big(\mathcal{D}[A]+\mathcal{D}[B]\big)\rho
    + O(\varepsilon^3), \\
    \mathcal{E}_A \circ \mathcal{E}_B(\rho)
    &=
    \rho
    + \varepsilon^2\big(\mathcal{D}[A]+\mathcal{D}[B]\big)\rho
    + O(\varepsilon^3).
\end{align}
Thus the two channels commute up to order $\varepsilon^2$, even when $[A,B]\neq 0$.

This approximate compatibility is sufficient for continuous monitoring. Taking a timestep $\varepsilon^2 \sim dt$ and iterating the channels yields, in the continuous time limit $dt \to 0$, an additive generator of the form
\begin{align}
    \frac{d\rho}{dt}
    =
    \mathcal{D}[A]\rho
    +
    \mathcal{D}[B]\rho
    + \cdots,
\end{align}
which is precisely the structure of the stochastic master equation used in the main text.

\medskip

\noindent
In summary, while sharp measurements of non-commuting observables are incompatible, weak measurements are compatible at the level required for continuous-time dynamics: their non-commutativity appears only at higher order in the measurement strength and vanishes in the stochastic limit.

\subsection{Alternative interpretation}
Furthermore, simultaneous monitoring of a complete operator basis may be understood, equivalently, as the limit of a rapid sequence of weak measurements performed along different measurement axes. In such a scheme, each observable is measured for a short interval $\delta t$, with the measurement basis cycled sufficiently quickly that no single observable dominates the dynamics \cite{liu2025measurementbasedquantumdiffusionmodels}.

In the limit $\delta t \to 0$, this sequential protocol generates an effective time-averaged evolution, in which the corresponding dissipators add linearly. This follows from a Trotter-type expansion of the composed channels \cite{trotter1959product}, and remains valid independently of whether the observables commute. Consequently, even beyond leading-order expansions in the measurement strength, rapidly interleaved weak measurements reproduce the same continuous-time stochastic master equation as simultaneous monitoring.

This provides an alternative operational interpretation of the dynamics considered in the main text, and further justifies the use of a complete set of (generally non-commuting) measurement channels.
\section{Derivation of the Bloch-space Langevin equation}
\label{app:LangevinDerivation}

In this appendix we derive the Bloch-coordinate stochastic differential equation quoted in Eq.~\eqref{eq:langevin_compact}. Starting from the monitored SME
\begin{align}
    d\rho
    =
    -i[H,\rho]\,dt
    +\sum_j \mathcal{D}[\xi_j]\rho\,dt
    +\sum_j \mathcal{D}[c_j]\rho\,dt
    +\sum_j \sqrt{\eta_j}\,\mathcal{H}[c_j]\rho\,dW_j,
\end{align}
with
\begin{align}
    H=h_j S_j,\qquad
    c_j=\sqrt{2k_j}\,S_j,\qquad
    \xi_j=\sqrt{2\gamma_j}\,S_j,
\end{align}
we project onto the $SU(d)$ basis using
\begin{align}
    dx_l=\Tr(S_l\,d\rho).
\end{align}
The Bloch decomposition is
\begin{align}
    \rho=\tfrac{\mathds{1}}{d}+\tfrac{1}{2} x_k S_k.
\end{align}

\subsection{Hamiltonian contribution}

Using \eqref{eq:SUd_algebra},
\begin{align}
    dx_l\big|_{H} & = -i \sum_j h_j \Tr \left([S_l,S_j]\rho\right) dt \nonumber\\
    & = \sum_{j,k} h_j f_{ljk} \Tr \left(S_k \rho\right) dt\nonumber\\
    & =  \sum_{j,k} h_j f_{ljk} x_k dt \label{eq:L1}.
\end{align}
Relabelling indices and using antisymmetry of $f_{ijk}$ gives
\begin{align}
    f^l_{H}=h_j f_{ljk}x_k,
\end{align}
as quoted in Eq.~\eqref{eq:hamiltonian_drift}.

\subsection{Dissipative contribution}

For a Hermitian jump operator $L=\sqrt{2\lambda}\,S_j$,
\begin{align}
    \mathcal D[L]\rho
    =
    2\lambda\left(S_j\rho S_j-\tfrac{1}{2}\{S_j^2,\rho\}\right).
\end{align}
The full dissipative contribution is therefore obtained by summing over all monitored and unmonitored channels with total rate
\begin{align}
    l_j=2k_j+2\gamma_j.
\end{align}
Thus
\begin{align}
    \sum_j \mathcal D[c_j]\rho+\sum_j \mathcal D[\xi_j]\rho
    =
    \sum_j l_j\Big(S_j\rho S_j-\tfrac{1}{2}\{S_j^2,\rho\}\Big).
\end{align}

To project this term, we repeatedly use
\begin{align}
    S_iS_j=\tfrac{2}{d}\delta_{ij}\mathbb I+\left(g_{ijk}+\tfrac{i}{2}f_{ijk}\right)S_k.
\end{align}
Expanding $S_j\rho S_j$ and $\{S_j^2,\rho\}$ in the Bloch basis and collecting the coefficients of $S_l$, one follows: 
\begin{align}
    dx_l\big|_{\mathcal{D}}  = &\sum_{j} l_{j} \text{Tr} \big(  \tfrac{1}{2}\left([S_l,S_j] + \{S_l,S_j\} \right)\rho S_j - \tfrac{1}{2} (\rho S_l + S_l \rho) 
    \tfrac{1}{2}\{S_j,S_j\} \big) dt \nonumber\\
    = &\sum_{j} \l_j \text{Tr} \bigg(  \Big\{ \tfrac{i}{2} \sum_m f_{ljm}S_m + \tfrac{2}{d}\mathds{1}_d\delta_{lj}+\sum_m g_{ljm}S_m \Big\} \rho S_j 
    - (\rho S_l + S_l \rho)\Big\{ \tfrac{1}{d}\mathds{1}_d + \tfrac{1}{2}\sum_m g_{jjm} S_m \Big\} \bigg) dt \\
     =& \sum_{j,m} l_j \text{Tr} \Bigl( \left(\tfrac{i}{2} f_{ljm} +g_{ljm}\right)\tfrac{1}{2} \left([S_j,S_m] + \{S_j,S_m\} \right)\rho \Bigr) dt \nonumber \\
    &  \hspace{9.0em} 
    + l_l \text{Tr}\left( \tfrac{2}{d} \mathds{1}_d S_l \rho \right)dt - \sum_j l_j \text{Tr}\left( \tfrac{2}{d} \mathds{1}_d S_l \rho \right)dt -\tfrac{1}{2} \sum_{j,m} l_j g_{jjm} \text{Tr} \left(\{S_l ,S_m\} \rho \right) dt\\
    =& \sum_{j,m} l_j \text{Tr} \Big( \big(\tfrac{i}{2} f_{ljm} +g_{ljm}\big) 
    \Big\{  \tfrac{2}{d}\mathds{1}_d\delta_{jm}+ \sum_{k} \tfrac{i}{2} f_{jmk}S_k + g_{jmk}S_k \Big\} \rho \Big) dt\nonumber\\
    &  \hspace{14.0em} 
    + x_l \tfrac{2}{d} \Big( l_l - \sum_j l_j\Big) dt  - \sum_{j,m} l_j g_{jjm} \text{Tr} \Big(\sum_k g_{lmk}S_k \rho \Big) dt \\
    =& \sum_{j,m} l_j \left(\tfrac{i}{2}f_{ljm} + g_{ljm}\right) \tfrac{2}{d} \delta_{jm} dt + x_l \tfrac{2}{d} \Big( l_l - \sum_j l_j\Big)  ~dt \nonumber\\
    &  \hspace{14.0em} 
    + \sum_{j,m,k} l_j \left( \left(\tfrac{i}{2} f_{ljm} +g_{ljm}\right) 
    \left\{ \tfrac{i}{2} f_{jmk} + g_{jmk} \right\} -  g_{jjm} g_{lmk} \right) x_k ~dt . \label{eq:Langevin2b}
\end{align}
Utilising the properties of the Kronecker delta, the term containing a $\delta_{jm}$ is summed over $m$. We can also rotate some indices for future simplification, and use repeated indices properties of the $SU(d)$ structure constant, $f_{ijj} =0$ valid for all $i$ and $j$ (from \eqref{eq:structure_identities}). 

In terms of the Langevin drift vector,
\begin{align}
    f^l_{\mathrm{diss}}
    =&
    \tfrac{2}{d}\sum_j l_j g_{ljj}
    +\tfrac{2}{d}x_l\!\Big(l_l-\sum_j l_j\Big) 
    +\sum_{j,m,k} l_j
    \Big[
        \left( \tfrac{i}{2}f_{jml}+g_{jml}\right)
        \left( \tfrac{i}{2}f_{jmk}+g_{jmk}\right)
        -g_{jjm}g_{lkm}
    \Big]x_k,
\end{align}
which is Eq.~\eqref{eq:dissipative_drift}.

\subsection{Stochastic contribution}

For the innovation superoperator with Hermitian jump operator $c_j=\sqrt{2k_j}\,S_j$,
\begin{align}
    \mathcal H[c_j]\rho
    =
    \sqrt{2k_j}\left(S_j\rho+\rho S_j-2\Tr(S_j\rho)\rho\right).
\end{align}
We find
\begin{align}
     \Tr\!\left(S_l\,\mathcal H[c_j]\rho\right)
    = & \sum_j \sqrt{2 k_j } \left[ \text{Tr}\left( \{S_l,S_j \} \rho \right) -2 x_l x_j \right] dW_j \\
     =& \sum_j \sqrt{2 k_j } \bigg[ \text{Tr}\Big( \Big\{ \tfrac{4}{d} \mathds{1}_d \delta_{lj} + 2 \sum_m g_{ljm}S_m \Big\} \rho \Big) -2 x_l x_j \bigg] dW_j \\
     =& \sqrt{2 k_l } ~\tfrac{4}{d} ~dW_l 
     +\sum_{j,m} 2\sqrt{2 k_j } ~g_{ljm} x_m ~dW_j  - \sum_j 2\sqrt{2 k_j }~  x_l x_j ~dW_j
     \label{eq:L3}
\end{align}

Including the efficiency factor $\sqrt{\eta_j}$ therefore gives the noise amplitudes
\begin{align}
    \sigma^l_j(\vec x)
    =
    \sqrt{2k_j\eta_j}
    \left(
        \tfrac{4}{d}\delta_{lj}
        +2g_{ljm}x_m
        -2x_lx_j
    \right),
\end{align}
which is Eq.~\eqref{eq:noise_amplitudes_main}.

Collecting all terms yields the Bloch-space Langevin equation
$dx_l=f^l(\vec x)\,dt+\sigma^l_j(\vec x)\,dW_j$,
with $f^l=f^l_{H}+f^l_{\mathrm{diss}}$.

\section{Diffusion tensor and drift vector simplifications}
\label{app:FPE_terms}
\subsection{Diffusion tensor and derivative identities}

In this appendix we record the explicit diffusion tensor and the Euclidean derivative identities used in the homogeneous calculations, under HS geometry.

Starting from the noise amplitudes
\begin{align}
    \sigma^\mu_j
    =
    \sqrt{2k_j\eta_j}
    \left(
        \tfrac{4}{d}\delta_{\mu j}
        +2g_{\mu jm}x_m
        -2x_\mu x_j
    \right),
\end{align}
the diffusion tensor is $D^{\mu\nu}= \sum_j \sigma^\mu_j\sigma^\nu_j$.
Assuming uncorrelated Wiener increments on different axes, $dW_j dW_i= \delta_{ij }dt$, expanding directly gives Eq.~\eqref{eq:Duv}, which we restate for convenience,
\begin{align}
    D^{\mu\nu}=  \sum_j 2 k_j \eta_j \Bigg[&
    \delta_{\mu j}\delta_{\nu j}\frac{16}{d^2}
    + \delta_{\mu j}\tfrac{8}{d}\sum_n g_{\nu j n}x_n
    - \delta_{\mu j}\tfrac{8}{d}x_\nu x_j
    + \delta_{\nu j}\tfrac{8}{d}\sum_m g_{\mu j m}x_m
    \nonumber\\
    &\hspace{0.2cm}
    + 4\sum_{m,n} g_{\mu jm}g_{\nu jn}x_n x_m
    - 4\sum_m g_{\mu jm}x_m x_\nu x_j
    - \delta_{\nu j}\tfrac{8}{d}x_\mu x_j
    - 4\sum_n g_{\nu jn}x_\mu x_j x_n
    + 4 x_\mu x_j x_\nu x_j
    \Bigg].
\end{align}

For homogeneous monitoring, $ k_j=k$, the diffusion simplifies to
\begin{align}\label{eq:app_homog_diffusionTensor}
    D^{\mu\nu}
    =
    2k\eta
    \Big[
        \delta_{\mu \nu} \frac{16}{d^2}&
        + \frac{16}{d} \sum_n g_{\nu \mu n}x_n
        + 4\sum_{j,m,n} g_{\mu jm} g_{\nu jn}x_n x_m
        - \frac{16}{d}x_\mu x_\nu 
        - 4\sum_{j,m}\big(g_{\mu j m }x_\nu + g_{\nu j m}x_\mu\big) x_m x_j + 4 x_j^2 x_\nu x_\mu
    \Big].
\end{align}
And for $d=2$, the homogeneous diffusion tensor simplifies to
\begin{align}\label{eq:app_d=2_diffusionTensor}
    D^{\mu\nu}
    =
    8k\eta \left[\delta_{\mu\nu} + (r^2-2)x_\mu x_\nu\right].
\end{align}

On the pure shell ($C=0$) for arbitrary dimensions, we can use identities from \eqref{eq:structure_identities} and \eqref{eq:pure_state_identities} to simplify the diffusion tensor expression~\eqref{eq:app_homog_diffusionTensor} further:
\begin{align}\label{eq:duDuv_C}
    D^{\mu\nu}
    =
    2k\eta
    \Big[
        \delta_{\mu \nu} \frac{16}{d^2}
        + \frac{16}{d} \sum_n g_{\nu \mu n}x_n
        + 4\sum_{j,m,n} g_{\mu jm} g_{\nu jn}x_n x_m
        - 8\left( 
        1-\tfrac{1}{d}\right) x_\mu x_\nu
    \Big],
\end{align}
which is Eq.~\eqref{eq:homo_Duv} for non-unit efficiency.

Next, we calculate the first derivative of $D^{\mu \nu}$
Differentiating the homogeneous tensor and simplifying with the structure-constant contractions yields 
\begin{align}\label{eq:duDuv}
    \partial_\mu D^{\mu\nu}
    =2k\eta \left[ 
    -16 d x_\nu + 4\frac{d^2-4}{d}x_\nu - 4(d^2 +2)\sum_{j,m}g_{\nu j m} x_m x_j + 4 (d^2+2) x_j^2 x_\nu
    \right],
\end{align}
where for $C=0$ yields $\partial_\mu D^{\mu\nu}= -8k\eta d\,x_\nu$, which is Eq.~\eqref{eq:divD_simplified}.

As used in evaluating \eqref{eq:iii'}, we now derive the trace of the diffusion tensor, $D^\mu_\mu$. Contracting indices for the homogeneous case
\begin{align}
    D^\mu_\mu & = 2k\eta \left[ 
    (d^2-1)\frac{16}{d^2}-\frac{16}{d}x_j^2 + 4\frac{d^2-4}{d}x_j^2-8\sum_{j,m,n}g_{jmn}x_j x_m x_n + 4x_j^2 x_i^2
    \right]
\end{align}
and for $d=2$ gives $D^{\mu}_{\mu}=8k\eta (3-2r^2+r^4)$.

When assessing steady state conditions, one requires the double divergence of the homogeneous diffusion tensor. A direct differentiation of Eq.~\eqref{eq:duDuv}, again using the identities of \eqref{eq:structure_identities}, yields
\begin{align}\label{eq:dvduDuv}
    \tfrac{1}{2}\partial_\nu\partial_\mu D^{\mu\nu}
    =
    4k\eta\bigg[
        (d^2+2)(d^2+1)\sum_j x_j^2
        -\frac{(3d^2+4)(d^2-1)}{d}
    \bigg],
\end{align}
which is Eq.~\eqref{eq:mixed_divdivD} in the main text.

For $d=2$, using $\sum_j x_j^2=1+2C$, this becomes
\begin{align}
    \tfrac{1}{2}\partial_\nu\partial_\mu D^{\mu\nu}
    =
    120k\eta (2C+1)-96k
    =
    240k\eta C+24k\eta.
\end{align}

\subsection{Drift vector and derivative identities}

We provide here some useful expressions for the drift vector $f^\nu$ and its derivatives under different conditions. We use \eqref{eq:structure_identities} throughout.

From \eqref{eq:hamiltonian_drift} and \eqref{eq:dissipative_drift}, the full drift vector can be defined, $f^l=f^l_{H}+f^l_{\mathrm{diss}}$. The Hamiltonian drift $f^l_{H}$ does not simplify under homogeneous conditions but, $f^l_{\mathrm{diss}}$ reduces to 
\begin{align}\label{eq:f_diss_homogeneous}
    f^l_{\mathrm{diss}} = -2dlx_l.
\end{align}

The derivative of $f^l_{H}$ vanishes---it is divergence free. The dissipative contribution yields $\partial_\nu f^\nu = -2d \sum_j l_j$, which in the homogeneous instance is equal to $2d(1-d^2)l$, where $l=2(k+\gamma)$.

We also use the expression $f^\nu \partial_\nu C$ where $\partial_\nu C = x_\nu$. Taking the product, we find $f^\nu_H \partial_\nu C =0$ by symmetry arguments and 
\begin{align}
    f^\nu_{\mathrm{diss}}\partial_\nu C
    =&
    \tfrac{2}{d}\sum_{j,\nu} l_j g_{\nu jj} x_\nu
    +\tfrac{2}{d}\sum_\nu x_\nu^2 \!\Big(l_\nu-\sum_j l_j\Big) 
    +\sum_{j,m,k,\nu} l_j
    \Big[
        \left(\tfrac{i}{2}f_{jm\nu}+g_{jm\nu}\right)
        \left(\tfrac{i}{2}f_{jmk}+g_{jmk}\right) 
        -g_{jjm}g_{\nu km}
    \Big]x_k x_\nu.
\end{align}
which does not simplify further. Under homogeneous rates, however this reduces dramatically to $f^\nu \partial_\nu C = -2ld\, |x|^2 $ which is exactly \eqref{eq:(i)fdC}.

\section{Derivation of the covariant Fokker--Planck equation}
\label{app:CovariantFPE}


We derive the covariant FPE associated with the stochastic dynamics introduced in Sec.~\ref{sec:FPE}. Consider the over-damped Langevin equation in It\^o stochastic differential equation form
\begin{align}
    dx^\nu
    =
    f^\nu(\vec{x},t)\,dt
    +
    \sigma^\nu_j(\vec{x},t)\,dW_j ,
    \label{eq:app_langevin}
\end{align}
where $f^\nu$ is the drift vector field and $\sigma^\nu_j$ are the noise amplitudes. The Wiener increments satisfy $dW_i\,dW_j=\delta_{ij}\,dt$ and $dW_j\,dt=0$. The associated diffusion tensor is $D^{\mu\nu}:=  \sum_j \sigma^\mu_j \sigma^\nu_j$.

Throughout, $\mathcal M$ denotes a Riemannian manifold with metric $g_{\mu\nu}$, volume element $dV=\sqrt{g}\,d^dx$ and covariant derivative $\nabla_\mu$. Inner products are written
\begin{align}
    \langle X,Y\rangle
    =
    g_{\mu\nu}X^\mu Y^\nu 
\end{align}
and let $\varrho(\vec x,t)$ denote the ensemble probability density with respect to $dV$, normalised as
\begin{align}
    \int_{\mathcal M} \varrho(\vec x,t)\,dV
    =
    1 .
\end{align}

\subsection{Continuity equation}

Probability conservation implies that, for any region
$\mathcal V\subset\mathcal M$ with boundary
$\partial\mathcal V$ and outward unit normal $\hat n$,
the rate of change of probability inside $\mathcal V$
equals the probability flux across the boundary:
\begin{align}
    \frac{d}{dt}
    \int_{\mathcal V}
    \varrho\,dV
    =
    -
    \int_{\partial\mathcal V}
    \langle J,\hat n\rangle\,dS ,
    \label{eq:prob_cons}
\end{align}
where $J^\mu$ is the probability current. Using the divergence theorem on $\mathcal M$,
\begin{align}
    \int_{\mathcal V}
    \nabla_\mu J^\mu\,dV
    =
    \int_{\partial\mathcal V}
    \langle J,\hat n\rangle\,dS ,
\end{align}
Eq.~\eqref{eq:prob_cons} becomes
\begin{align}
    \int_{\mathcal V}
    \left(
        \partial_t\varrho
        +
        \nabla_\mu J^\mu
    \right)dV
    =
    0 .
\end{align}
Since $\mathcal V$ is arbitrary,
\begin{align}\label{eq:app_conservationFPE}
    \partial_t\varrho
    +
    \nabla_\mu J^\mu
    =
    0 ,
\end{align}
which is the covariant continuity equation.

\subsection{Applying It\^o's Lemma}

To determine the probability current, let $\kappa(\vec x)$ be a smooth test function. Applying It\^o's lemma to Eq.~\eqref{eq:app_langevin} in local coordinates gives
\begin{align}
    d\kappa
    =
    \partial_\nu\kappa\,dx^\nu
    +
    \frac12
    \partial_\mu\partial_\nu\kappa\,
    dx^\mu dx^\nu .
\end{align}
The second partial derivative is not tensorial and therefore cannot appear directly in a coordinate-invariant expression. Since $\partial_\nu\kappa$ is a covector, its covariant derivative is
\begin{align}
    \nabla_\mu(\partial_\nu\kappa)
    =
    \partial_\mu\partial_\nu\kappa
    -
    \Gamma^\lambda_{\mu\nu}
    \partial_\lambda\kappa ,
\end{align}
where $\Gamma^\lambda_{\mu\nu}$ are the Christoffel symbols associated with the metric $g_{\mu\nu}$. Rearranging,
\begin{align}
    \partial_\mu\partial_\nu\kappa
    =
    \nabla_\mu(\partial_\nu\kappa)
    +
    \Gamma^\lambda_{\mu\nu}
    \partial_\lambda\kappa .
\end{align}
Using $dx^\mu dx^\nu=D^{\mu\nu}dt$, It\^o's lemma becomes
\begin{align}
    d\kappa
    =
    \left[
        h^\lambda\partial_\lambda\kappa
        +
        \frac12
        D^{\mu\nu}
        \nabla_\mu(\partial_\nu\kappa)
    \right]dt
    +
    \sigma^\nu_j
    \partial_\nu\kappa\,dW_j ,
\end{align}
where the effective drift is
\begin{align}
    h^\lambda
    :=
    f^\lambda
    +
    \frac12
    \Gamma^\lambda_{\mu\nu}
    D^{\mu\nu}.
\end{align}

Averaging over the stochastic noise eliminates the $dW_j$ contribution, giving
\begin{align}
    \frac{d}{dt}
    \langle\!\langle\kappa\rangle\!\rangle
    =
    \left\langle\!\!\!\left\langle
        h^\lambda\partial_\lambda\kappa
        +
        \frac12
        D^{\mu\nu}
        \nabla_\mu(\partial_\nu\kappa)
    \right\rangle\!\!\!\right\rangle .
\end{align}
The backward generator therefore takes the form
\begin{align}
    L\kappa
    =
    h^\lambda\partial_\lambda\kappa
    +
    \frac12
    D^{\mu\nu}
    \nabla_\mu(\partial_\nu\kappa),
\end{align}
and integrating over the manifold gives
\begin{align}\label{eq:backward_generator_integral}
    \frac{d}{dt}
    \int_{\mathcal M}
    \kappa\,\varrho\,dV
    =
    \int_{\mathcal M}
    (L\kappa)\,\varrho\,dV ,
\end{align}
where we have dropped the $\langle\!\langle\cdots\rangle\!\rangle$ notation and, from now on, work exclusively with noise-averaged objects.

\subsection{Integration by parts}

Using the covariant divergence theorem over the entire manifold $\mathcal{M}$,
\begin{align}
    \int_{\mathcal M}
    \nabla_\mu V^\mu\,dV
    =
    0 ,
\end{align}
for sufficiently regular vector fields satisfying appropriate boundary conditions, we now integrate by parts to move derivatives from $\kappa$ onto $\varrho$. For the drift term,
\begin{align}
    \int_{\mathcal M}
    h^\nu(\partial_\nu\kappa)\,\varrho\,dV
    &=
    \int_{\mathcal M}
    \nabla_\nu(\kappa h^\nu\varrho)\,dV
    -
    \int_{\mathcal M}
    \kappa\,\nabla_\nu(h^\nu\varrho)\,dV
    \nonumber\\
    &=
    -
    \int_{\mathcal M}
    \kappa\,\nabla_\nu(h^\nu\varrho)\,dV ,
\end{align}
where the boundary contribution vanishes. For the diffusion term, integrating by parts twice gives
\begin{align}
    \int_{\mathcal M}
    D^{\mu\nu}
    \nabla_\mu(\partial_\nu\kappa)
    \,\varrho\,dV
    &=
    -\int_{\mathcal M}
    (\partial_\nu\kappa)
    \nabla_\mu(D^{\mu\nu}\varrho)\,dV
    \nonumber\\
    &=
    \int_{\mathcal M}
    \kappa\,
    \nabla_\nu\nabla_\mu
    (D^{\mu\nu}\varrho)\,dV ,
\end{align}
again assuming that the boundary terms vanish.

Combining these two contributions into Eq.~\eqref{eq:backward_generator_integral} yields
\begin{align}
    \frac{d}{dt}
    \int_{\mathcal M}
    \kappa\,\varrho\,dV
    =
    \int_{\mathcal M}
    \kappa
    \left[
        -\nabla_\nu(h^\nu\varrho)
        +
        \frac12
        \nabla_\nu\nabla_\mu
        (D^{\mu\nu}\varrho)
    \right]dV .
\end{align}
Lastly, since $\kappa$ is arbitrary, the integrands must agree pointwise, giving
\begin{align}
    \partial_t\varrho
    =
    -\nabla_\nu(h^\nu\varrho)
    +
    \frac12
    \nabla_\nu\nabla_\mu
    (D^{\mu\nu}\varrho).
    \label{eq:FPE_avg}
\end{align}
Equivalently, this may be written in conservation-law form as Eq.~\eqref{eq:app_conservationFPE}, with probability current
\begin{align}
    J^\nu
    =
    h^\nu\varrho
    -
    \frac12
    \nabla_\mu(D^{\mu\nu}\varrho).
\end{align}

\section{Pure-state identities in Bloch coordinates}
\label{app:PureStateIdentities}

In this appendix we derive the identities used repeatedly on the pure-state manifold.

A state is pure if and only if
\begin{align}
    \rho^2=\rho.
\end{align}
Substituting the Bloch decomposition
\begin{align}
    \rho=\tfrac{\mathbb I}{d}+\tfrac{1}{2} x_a S_a
\end{align}
gives
\begin{align}
    \rho^2
    =
    \frac{\mathbb I}{d^2}
    +\frac1d x_a S_a
    +\frac{1}{4} x_a x_b S_aS_b.
\end{align}
Using
\begin{align}
    S_aS_b
    =
    \tfrac{2}{d}\delta_{ab}\mathbb I
    +\left(g_{abn}+\tfrac{i}{2}f_{abn}\right)S_n,
\end{align}
we find
\begin{align}
    \rho^2
    =&
    \left(
        \frac{1}{d^2}
        +\frac{1}{2d}x_ax_a
    \right)\mathbb I
    +\left(
        \frac{1}{d}x_n
        +\frac{1}{4} x_ax_b g_{abn}
    \right)S_n,
\end{align}
since the antisymmetric $f_{abn}$ term vanishes upon contraction with the symmetric product $x_ax_b$.

Matching the identity component of $\rho^2=\rho$ gives
\begin{align}
    \frac{1}{d^2}+\frac{1}{2d}x_ax_a=\frac1d,
\end{align}
hence
\begin{align}
    x_ax_a=\frac{2(d-1)}{d}.
\end{align}
Matching the generator component gives
\begin{align}
    \frac1d x_n+\tfrac{1}{4} x_ax_b g_{abn}
    =
    \tfrac{1}{2} x_n,
\end{align}
so that
\begin{align}
    x_ax_b g_{abn}
    =
    \frac{2(d-2)}{d}x_n.
\end{align}
These are precisely the identities quoted in Eq.~\eqref{eq:pure_state_identities}.

\subsection{Tangential derivative identity}

A further identity used in the proof of condition \eqref{eq:cond_iv} is obtained by differentiating
\begin{align}
    g_{abn}x_ax_b=\frac{2(d-2)}{d}x_n
\end{align}
along a tangent direction $\tilde v_a$. Since $g_{abn}$ is symmetric in $a,b$,
\begin{align}
    2g_{abn}\tilde v_a x_b
    =
    \frac{2(d-2)}{d}\tilde v_n,
\end{align}
and therefore
\begin{align}
    g_{abn}\tilde v_a x_b
    =
    \frac{d-2}{d}\tilde v_n,
\end{align}
which is Eq.~\eqref{eq:tangent_g_identity}.

\section{Wasserstein calculation}
\label{app:wasserstein_calc}

To quantify convergence to the desired mixed-state ensemble, we compute an entropically regularised optimal-transport distance between the generated ensemble and a reference point cloud sampled from the target measure---referred to in-text as the Wasserstein distance. Take the empirical measures
\begin{align}
    A
    =
    \sum_m p_m\delta_{\mathbf{x}_m},
    \qquad
    B
    =
    \sum_n q_n\delta_{\mathbf{y}_n},
\end{align}
where $\mathbf{x}_m$ and $\mathbf{y}_n$ denote sample points in state space, $\delta_{\mathbf{x}}$ is a Dirac delta measure concentrated at $\mathbf{x}$, and $\{p_m\}$ and $\{q_n\}$ are non-negative weights satisfying $\sum_m p_m=\sum_n q_n=1$.
The cost matrix associated with the Wasserstein-1 distance is then defined by the absolute distances
\begin{align}
    C_{mn}=|\mathbf{x}_m-\mathbf{y}_n|.
\end{align}
Next, we compute the Sinkhorn-regularised transport cost
\begin{align}\label{eq:app_WasstersteinCalc}
    W_\varepsilon(A,B)
    =
    \min_{\Pi}
    \sum_{mn}\Pi_{mn}C_{mn}
    +
    \varepsilon
    \sum_{mn}\Pi_{mn}
    \left(
        \log\Pi_{mn}-1
    \right),
\end{align}
which is a variational quantity, found via numeric optimisation, and is subject to the formulation
\begin{align}
    \sum_n\Pi_{mn}=p_m,
    \qquad
    \sum_m\Pi_{mn}=q_n .
\end{align}
The second term in Eq.~\eqref{eq:app_WasstersteinCalc} is an entropic regularisation controlled by the parameter $\varepsilon$. It renders the optimisation problem strictly convex and computationally efficient, while recovering the unregularised optimal-transport problem in the limit $\varepsilon\to0$ \cite{Cuturi2013}. The quantity
$
-\sum_{mn}\Pi_{mn}\log\Pi_{mn}
$
is the Shannon entropy of the transport plan, so the regularisation favours smoother couplings distributed over many transport paths (with larger entropy), rather than highly concentrated assignments.

For pure states, as in Fig.~\ref{fig:pure_state_wasserstein}, the reference distribution is constructed by sampling randomly from the uniform Bloch Sphere.
For the HS mixed state benchmark in Figs.~\ref{fig:mixed_state_wasserstein} and \ref{fig:postmix_hs_wasserstein_weights}, the reference cloud is sampled uniformly from the Bloch ball. For the Bures benchmark in Fig.~\ref{fig:postmix_bures_wasserstein_weights}, the angular coordinates are sampled uniformly while the radial coordinate is drawn from Eq.~\eqref{eq:Bures_radial}. The Wasserstein curves shown in the main text are obtained by repeating this calculation for each saved time point along the evolution.

\section{Failure of fixed state-independent approach}
\label{app:state-independent_failure}

In this appendix we provide a brief calculation that demonstrates the impossibility of a single choice of $\eta$ and $\gamma$ generating the HS uniform density over the full Bloch body. This is achieved by examining the steady state condition for potential dynamics under a fixed choice of $\eta$ and $\gamma$. 

Presupposing homogeneity (intuitively required to build an isotropic distribution), and starting from the equation for $\partial_\nu f^\nu = -4d(d^2-1)(k+\gamma)$ provided above \eqref{eq:f_diss_homogeneous} and the diffusion tensor from \eqref{eq:dvduDuv}, we get
\begin{align}
    \partial_\nu f^\nu -\tfrac{1}{2} \partial_\nu \partial\mu D^{\mu \nu} = -4d(d^2-1)(k+\gamma) 
    + 4k\eta \bigg[ 
    \frac{(3d^2+4)(d^2-1)}{d}-(d^2+2)(d^2+1)x_j^2
    \bigg].
\end{align}
This expression must vanish if the uniform state is a steady state. This is only the case if
\begin{align}
    \frac{(d^2-1)[ k\eta(3d^2+4) - d^2(k+\gamma)]}{k\eta d (d^2+2)(d^2+1)} = x_j^2 \quad \forall j.
\end{align}
Given $x_j^2$ varies across the Bloch radius (between $0$ and $1$), it is apparent that no combination of parameters $k$, $\eta$ and $\gamma$ on the LHS can satisfy the above equation, unless they themselves are functions $x_j$. This motivates our decision to investigate state-dependent parameters in generating random mixed states.


\section{Derivation of the ellipticity condition}
\label{app:III_derivation}

In this appendix we derive the operator form of the kernel condition used in the validation of \eqref{eq:III} in Sec.~\ref{sec:MixedStateProofs}. Starting from the homogeneous noise amplitudes \eqref{eq:noise_amplitudes_main}, the condition $
    v_\mu \sigma^\mu_j = 0~ \forall j $
is equivalent to
\begin{align}
    \tfrac{2}{d} v_j + g_{\mu j n} x_n v_\mu - (x \cdot v)\, x_j = 0
    \qquad \forall j,
    \label{eq:app_component_condition}
\end{align}
where $x \cdot v := x^\mu v_\mu$.

We now express this condition in operator form, defining $ V := \tfrac{1}{2} v_\mu S_\mu$
so that $V$ is Hermitian and traceless. Using the Bloch representation of $\rho$ from \eqref{eq:bloch_rep}, we compute the anticommutator
\begin{align}
    \{\rho, V\}
    =
    \tfrac{1}{d} V
    + \tfrac{1}{4} x_n v_\mu \{S_n, S_\mu\}.
\end{align}
Using the algebra \eqref{eq:SUd_algebra}, this becomes
\begin{align}
    \{\rho, V\}
    =
    \frac{x \cdot v}{d} \mathds{1}
    +
    \left(
        \frac{1}{d} v_j + \frac{1}{2} g_{\mu j n} x_n v_\mu
    \right) S_j.
\end{align}
Meanwhile, the expectation value of $V$ is
\begin{align}
    \Tr(\rho V)
    =
    \tfrac{1}{2} x_\mu v_\mu
    =
    \tfrac{1}{2} (x \cdot v).
\end{align}
Therefore,
\begin{align}
    2 \Tr(\rho V)\, \rho
    =
    \frac{x \cdot v}{d} \mathds{1}
    +
    \frac{1}{2} (x \cdot v)\, x_j S_j.
\end{align}
Subtracting, we obtain
\begin{align}
    \{\rho, V\} - 2 \Tr(\rho V)\, \rho
    =
    \tfrac{1}{2}
    \left(
        \tfrac{2}{d} v_j
        + g_{\mu j n} x_n v_\mu
        - (x \cdot v) x_j
    \right) S_j.
\end{align}

Since the generators $S_j$ are linearly independent, the vanishing of this expression is equivalent to Eq.~\eqref{eq:app_component_condition}. Thus the kernel condition is equivalent to
\begin{align}
    \{\rho, V\} = 2 \Tr(\rho V)\, \rho.
\end{align}
This establishes Eq.~\eqref{eq:III_operator_condition} used in the main text.

\section{Bures metric calculations for qubits}
\label{app:BuresDetails}

In this appendix we summarise the derivation of the Bures metric expressions used in Sec.~\ref{sec:MixedStateProofs}, leading to the solution for the state-dependent rates that produce the uniform ensemble.

\subsection{Metric and Christoffel symbols}

The standard radial form of the Bures metric for a qubit is \cite{BengtssonZyczkowski,HUBNER1992239}
\begin{align}\label{eq:app_bures_ds2}
    ds^2 = \frac{dr^2}{1-r^2} + r^2 d\Omega^2
\end{align}
where $r^2=x^2+y^2+z^2$ and $d\Omega^2 = d\theta^2 + \sin ^2 (\theta) \,d\varphi^2$ with $\theta$ and $\varphi$ the usual polar and azimuthal angle respectively. Since the FPE is formulated in Cartesian Bloch coordinates, we express the metric in the same coordinate system. Using
\begin{align}
    dr^2 = dx^2 + dy^2 + dz^2 - \frac{(x\,dx+y\,dy+z\,dz)^2}{r^2},
\end{align}
one obtains the line element
\begin{align}
    ds^2 = dx^2 + dy^2 + dz^2 + \frac{(x\,dx+y\,dy+z\,dz)^2}{1-r^2}.
\end{align}
In Cartesian Bloch coordinates, using the fact that $ds^2 = m_{\mu \nu}dx^\mu dx^\nu$, this corresponds to the metric 
\begin{align}\label{eq:app_muv}
    m_{\mu\nu} = \delta_{\mu\nu} + \frac{x_\mu x_\nu}{1-r^2}.
\end{align}
The inverse metric is
\begin{align}
m^{\mu\nu}
=
\delta^{\mu\nu}
-
x^\mu x^\nu.
\end{align}

We now recall the definition of the Levi--Civita connection,
\begin{align}
    \Gamma^k_{\mu\nu}
    =
    \tfrac{1}{2} m^{k\lambda}
    \left(
        \partial_\mu m_{\nu\lambda}
        +\partial_\nu m_{\mu\lambda}
        -\partial_\lambda m_{\mu\nu}
    \right),
\end{align}
and compute the derivative of Eq.~\eqref{eq:app_muv},
\begin{align}
    \partial_\mu m_{\nu\lambda}
    =
    \frac{\delta_{\mu\nu}x_\lambda + \delta_{\mu\lambda}x_\nu}{1-r^2}
    + \frac{2x_\mu x_\nu x_\lambda}{(1-r^2)^2}.
\end{align}
After simplification, we find a particularly simple form for the Levi--Civita connection,
\begin{align} \label{eq:app_BuresChristoffel}
    \Gamma^k_{\mu\nu} = x^k m_{\mu\nu}.
\end{align}

\subsection{Covariant derivatives of the drift vector and diffusion tensor}

The stationarity condition depends on the divergence of the effective drift, $h^\nu =
f^\nu
+
\frac12
\Gamma^\nu_{ij}D^{ij}$, which we evaluate term by term.
The covariant divergence of a vector field $F^\nu$ is
\begin{align}
    \nabla_\nu F^\nu = \partial_\nu F^\nu + \Gamma^\nu_{\nu k} F^k,
\end{align}
where, for the Bures metric, $\Gamma^\nu_{\nu k}=\frac{x_k}{1-r^2}$.
Substituting the homogeneous deterministic drift from Eq.~\eqref{eq:mixed_drift} and simplifying yields
\begin{align}
    \nabla_\nu f^\nu
    =
    l(C)\left[-12+\frac{2(1+2C)}{C}\right]
    -4(1+2C)l'(C).
\end{align}
The geometric contribution to the effective drift is, using Eq.~\eqref{eq:app_BuresChristoffel}, 
\begin{align}
    \tfrac12 \Gamma^\nu_{ij} D^{ij} = 4kx^\nu (3-r^2)
\end{align}
and its divergence in terms of $C$,
\begin{align}
    \nabla_\nu \big( \tfrac12 \Gamma^\nu_{ij} D^{ij}\big) 
    &= 4k\big(3-8C-\tfrac1C \big).
\end{align}

Finally, starting from Eq.~\eqref{eq:app_d=2_diffusionTensor}, $
    D^{\mu\nu}
    =
    8k\left[\delta_{\mu\nu} + (r^2-2)x_\mu x_\nu\right]$, 
we aim to compute
\begin{align}
    \nabla_\mu D^{\mu\nu}
    =
    \partial_\mu D^{\mu\nu}
    +\Gamma^\mu_{\mu k} D^{k\nu}
    +\Gamma^\nu_{\mu k} D^{\mu k}.
\end{align}
Using Eq.~\eqref{eq:app_BuresChristoffel} and simplifying term by term gives
\begin{align}
    \nabla_\mu D^{\mu\nu}
    =
    -32k(1-r^2)x^\nu.
\end{align}
Applying a second covariant derivative yields
\begin{align}
    \nabla_\nu \nabla_\mu D^{\mu\nu}
    =
    32k(4r^2-3).
\end{align}

\subsection{ODE for $\gamma(C)$}

Substituting the above expressions into the stationarity condition, $\nabla_\nu h^\nu - \tfrac{1}{2}\nabla_\nu \nabla_\mu D^{\mu\nu} = 0$, and expressing $r^2=1+2C$, gives the differential equation for $\gamma_B (C)$ as in Eq.~\eqref{eq:ODE_BuresGamma}. Motivated by the linear dependence of the inhomogeneous term on $C$, we seek a solution of the form $\gamma_B(C)=bC$. Substituting this ansatz into the ODE immediately yields Eq.~\eqref{eq:Gamma_Bures_solution}.


\section{Radial densities for qubit metrics}
\label{app:radial_densities}

For a qubit, any unitarily invariant measure depends only on the Bloch radius $r$ and therefore separates into an angular Haar part and a radial density. Thus the desired probability density over radii can be written as
\begin{align}
    p_*(r)\,dr
    =
    \frac{\Omega(r)\,dr}{\int_0^1\Omega(r')\,dr'} ,
\end{align}
where $\Omega(r)\,dr$ is the volume contained within the spherical shell $[r,r+dr]$. For the HS metric, the Bloch ball is Euclidean and
\begin{align}
    \Omega_{\mathrm{HS}}(r)=4\pi r^2 .
\end{align}
Normalising over the unit ball gives
\begin{align}
    p_{\mathrm{HS}}(r)=3r^2 .
\end{align}

For the Bures metric, the volume element induced by the metric from Eq.~\eqref{eq:app_muv} is proportional to $\sqrt{\det m}\,d^3x $,
so that the qubit volume element contains an additional boundary factor,
\begin{align}
    \Omega_{\mathrm{B}}(r)
    \propto
    \frac{r^2}{\sqrt{1-r^2}} .
\end{align}
Using $\int_0^1 \frac{r^2}{\sqrt{1-r^2}}\,dr=  \frac{\pi}{4}$,
one obtains
\begin{align}
    p_{\mathrm{B}}(r)
    =
    \frac{4}{\pi}
    \frac{r^2}{\sqrt{1-r^2}} .
\end{align}
This integrable divergence at $r=1$ reflects the strong boundary weighting of the Bures measure. Within the state-dependent-rate construction, reproducing this behaviour leads to the regularity and positivity difficulties discussed in Sec.~\ref{sec:MixedStateProofs}, whereas the post-mixing method can reproduce it through increased weight near $\eta=1$.

\section{Post-mixing of $\gamma$}
\label{app:Gamma_postmixing}

In this appendix we briefly repeat the analysis of Sec.~\ref{sec:inefficient} for fixed measurement efficiency $\eta=1$ and varying fixed decoherence rate $\gamma$. In parallel to Eq.~\eqref{eq:inefficieny_SME}, we consider the isotropic SME with additional decoherence
\begin{align}
    d\rho
    =
    2(k+\gamma)\sum_{j}\mathcal{D}[S_j]\rho\,dt
    +
    \sqrt{2 k}\sum_{j}\mathcal{H}[S_j]\rho\,dW_j .
\end{align}

For zero Hamiltonian and working in the HS geometry, we get the detailed balance property \eqref{eq:i'}
\begin{align}
    f^\nu - \tfrac{1}{2}\partial_\mu D^{\mu \nu} = &-4 d (k+\gamma) x_\nu+ k \bigg[
    \frac{4(3d^2+4)}{d} x_\nu 
     + 4(d^2+2) \sum_{j,m}g_{\nu jm}x_m  x_j   
     - 4 (d^2+2) x_j^2 x_\nu 
    \bigg].
\end{align}
In the qubit case, the detailed-balance expression vanishes at the radius $r_\gamma$ which obeys
\begin{align}
    r_\gamma^2
    =1-
    \frac{\gamma}{3k},
    \qquad
    \gamma \in\left[0,3k\right],
    \label{eq:rGamma_app}
\end{align}
so that $\gamma=3k$ corresponds to the maximally mixed state and $\gamma=0$ to the pure-state boundary. So, as is the case for inefficient measurements, for a single fixed value of $\gamma$, the detailed balance condition predicts a preferred radius that can be tuned over the full interior of the Bloch ball.

Property \eqref{eq:ii'} follows immediately from the validation of condition \eqref{eq:III} in the main text. The moment-balance radius property \eqref{eq:iii'} for fixed decoherence rates reduces to
\begin{align}
    \tfrac{1}{8}Lr^2 = 3k - 2(2k+\gamma)r^2 + kr^4
\end{align}
that can be compared with Eq.~\eqref{eq:iii'_evaluated} via division with $k$. Setting the left-hand side equal to zero defines the moment-balance radius $r_L$. Using the quadratic formula for the solution of $r_{L,\pm}^2$, we get
\begin{align}
    r_{L,\pm}^2
    =
    2+\frac{\gamma}{k}
    \pm
    \sqrt{
        \left(2+\frac{\gamma}{k}\right)^2-3
    } .
\end{align}
The plus branch lies outside the Bloch ball, so the physical solution is given by the negative branch. Equivalently, one can solve for $\gamma$ in terms of $r_L^2$, finding a positive rate for every $0<r_L^2<1$, with $\gamma=0$ giving $r_L^2=1$ and $\gamma\to\infty$ giving $r_L^2\to0$. Thus, as with efficiency sampling, varying a fixed decoherence rate can in principle tune the moment-balance radius across the full Bloch-ball interior.

The detailed-balance radius $r_\gamma$ in Eq.~\eqref{eq:rGamma_app} is physical only for $0\leq\gamma\leq3k$, whereas the moment-balance radius remains physical for all $\gamma\geq0$. This suggests that post-mixing over fixed decoherence rates should also provide a viable family of radial kernels. We focus on efficiency post-mixing in the main text, but the same construction may be implemented by replacing $P(r|\eta)$ with $P(r|\gamma)$.


\end{document}